\documentclass[aps,prd,superscriptaddress,nofootinbib,groupedaddress,amsfonts,floatfix,twocolumn]{revtex4}
\usepackage{graphicx,amsmath,amssymb,amstext,amssymb,amsbsy,amsfonts,amsthm,lscape,units}

\newcommand{\lp}{\left(}
\newcommand{\rp}{\right)}
\newcommand{\lb}{\left[}
\newcommand{\rb}{\right]}

\newcommand{\pr}{^{\prime}}
\newcommand{\prpr}{^{\prime \prime}}

\newcommand{\beq}{\begin{equation}}
\newcommand{\eeq}{\end{equation}}

\newcommand{\mcl}{\mathcal{L}}
\newcommand{\mch}{\mathcal{H}}

\newcommand{\mct}{\mathcal{T}}

\newcommand{\Upc}{U_{+c}} 
\newcommand{\Ups}{U_{+s}} 
\newcommand{\Umc}{U_{-c}} 
\newcommand{\Ums}{U_{-s}}

\usepackage{color}

\begin{document}

\title{Relic Cosmological Vector Fields and Inflationary Gravitational Waves}

\newcommand{\Dartmouth}{Department of Physics and Astronomy, Dartmouth College, 6127 Wilder Laboratory, Hanover, NH 03755}

\author{Avery~J.~Tishue}
\email{avery.tishue.gr@dartmouth.edu}
\author{Robert~R.~Caldwell}
\affiliation{\Dartmouth}

\begin{abstract}
We show that relic vector fields can significantly impact a spectrum of primordial gravitational waves in the post-inflationary era. We consider a triplet of U(1) fields in a homogeneous, isotropic configuration. The interaction between the gravitational waves and the vector fields, from the end of reheating to the present day, yields novel spectral features. The amplitude, tilt, shape, and net chirality of the gravitational wave spectrum are shown to depend on the abundance of the electric- and magnetic-like vector fields. Our results show that even a modest abundance can have strong implications for efforts to detect the imprint of gravitational waves on the cosmic microwave background polarization. We find that a vector field comprising less than $2\%$ of the energy density during the radiation dominated era can have a greater than order unity effect on the predicted inflationary gravitational wave spectrum.

\end{abstract}

\maketitle

\section{Introduction}
\label{sec:sec1}

The inflationary paradigm is widely regarded as the leading framework to describe the early stages of the Hot Big Bang cosmology \cite{Guth:1980zm,Linde:1981mu,Albrecht:1982wi}. Indeed, the inflationary prediction of a nearly scale-invariant spectrum of primordial adiabatic density perturbations \cite{Mukhanov:1981xt,Hawking:1982cz,Guth:1982ec,Starobinsky:1982ee,Bardeen:1983qw} underpins the successes of the Standard Cosmological Model. Inflation also predicts a nearly scale-invariant spectrum of primordial gravitational waves (GWs) \cite{Grishchuk:1974ny,Starobinsky:1979ty,Rubakov:1982df,Fabbri:1983us,Abbott:1984fp}. Detection of the unique signature of these long wavelength GWs in the polarization pattern of the cosmic microwave background (CMB) would provide strong confirmation of inflation \cite{Kamionkowski:1996zd,Seljak:1996gy}. Consequently, there is an enormous experimental effort to detect the polarization signal \cite{Ade:2017uvt,Ade:2018gkx,Ade:2018sbj,Akrami:2018odb,Tristram:2020wbi,Abazajian:2016yjj,Hazumi:2019lys}. Similarly, there is a concerted theoretical effort to connect such a signal to the physics of inflation \cite{Spergel:1997vq,Kamionkowski:1997av,Knox:2002pe,Baumann:2014cja,Kamionkowski:2015yta}. Yet, despite these successes and prospects, there is no leading theoretical model of inflation, and investigations into the fundamental physics origin of the inflationary era is an ongoing concern. Most efforts focus on building a sufficiently flat potential for slow-roll scalar field evolution. However, Planck-scale quantum corrections make it a challenge to obtain viable inflation with an appreciable level of GWs \cite{Lyth:1996im}.

A new approach, consisting of a method to achieve slow roll inflation in a steep potential, has recently garnered attention. In this scenario, a coupling between an axion-like inflaton and the vacuum expectation value (vev) of a vector gauge field, through the Chern-Simons term, acts like a brake on the inflaton evolution \cite{Anber:2009ua,Adshead:2012kp}. These scenarios can produce sufficient inflation without resorting to fine tuning or Planck-scale physics. Variations on the basic model have been explored \cite{Maleknejad:2011jw,Maleknejad:2012fw,Noorbala:2012fh,Adshead:2013nka,Adshead:2016omu,Adshead:2017hnc,Dimastrogiovanni:2018xnn,Domcke:2018rvv,Domcke:2019mnd}, including the impact on reheating \cite{Adshead:2015pva,Adshead:2017xll,Adshead:2018doq,Adshead:2019igv,Adshead:2019lbr}. Models such as chromo-natural inflation \cite{Adshead:2012kp} organize the gauge field under SU(2) to maintain isotropy and spatial homogeneity, though other group structures are feasible. But all such models share a common feature: the tensor shear due to the gauge field vev strongly affects GW amplification and evolution during inflation \cite{Anber:2012du,Adshead:2013qp,Namba:2013kia,Maleknejad:2016qjz,Maleknejad:2016dve,Dimastrogiovanni:2016fuu,Fujita:2017jwq,Caldwell:2017chz,Agrawal:2017awz,Agrawal:2018mrg,Fujita:2018ndp,Watanabe:2020ctz}. The upshot is that these models overwrite the standard inflationary expressions that connect the primordial tensor amplitude to the scale of inflation or the tensor tilt to the slow roll parameters. A significant piece of this story is missing, however; the gauge field vev may be expected to survive inflation and reheating \cite{Adshead:2015pva,Adshead:2017xll,Adshead:2019lbr}, and further transform the GW spectrum.

There are various scenarios in which relic vector fields emerge from 
inflation \cite{Davis:2000zp,Dimopoulos:2001wx,Demozzi:2009fu,Adshead:2016iae,Kandus:2010nw}. In the simplest cases, a coupling between the inflaton or spectator scalar and a U(1) field pumps energy into the vector field, producing a spectrum of long-wavelength vector perturbations. 
The cumulative effect of superhorizon modes
contributes an effective homogeneous background, consisting of both electric and magnetic components \cite{Anber:2009ua,Domcke:2018rvv}.

In this paper, we investigate the effect of the relic vector fields on the primordial GW spectrum during the post-inflationary era. For simplicity, we consider a toy model consisting of a triplet of U(1) fields with vevs that are arranged in a spatially homogeneous and isotropic configuration. Such a configuration may occur as a central feature of inflation with Abelian fields \cite{Anber:2009ua} but may also arise from non-Abelian gauge field inflation, in the weakly-coupled regime \cite{Domcke:2018rvv}, or as a consequence of the post-inflationary thermalization process \cite{Davis:2000zp,Dimopoulos:2001wx,Adshead:2015pva}. The GW - gauge field system is parameterized by the amplitude of the electric- and magnetic-like field strengths, with constant fractional energy density during the radiation era. We show that the spectrum amplitude, tilt, shape, and net chirality are transformed. This is a new twist on cosmic archaeology \cite{Cui:2017ufi}: the imprinting of a non-standard expansion history \cite{Allahverdi:2020bys} or particle physics history \cite{Seto:2003kc,Boyle:2005se,Watanabe:2006qe,Nakayama:2008ip,Kuroyanagi:2011fy,Saikawa:2018rcs,Caldwell:2018giq,Figueroa:2019paj} on a GW spectrum. Through examples, we illustrate the consequences for CMB probes of primordial B-modes and efforts to directly detect a stochastic GW background. In the context of these scenarios, the strong amplification or suppression of a primordial GW spectrum has significant observational consequences.
 
The paper is organized as follows. In Sec.~\ref{sec:model} we introduce the model and conventions. We present our results in Sec.~\ref{sec:Results} and interpretations in Sec.~\ref{discussion}. Calculation details are presented in the appendices.

\section{Model and Conventions}
\label{sec:model} 

The vector field sector of this model is composed of a triplet of classical U(1) fields, $A_\mu ^{(i)}$, with each copy indexed by $i=1-3$.
Hence, after the relaxation and decay of the inflaton through the epoch of reheating, all remaining particles and fields are given by the Lagrangian density 
\beq
\mcl = \frac{1}{2}M_P^2 R - \frac{1}{4}F_{\mu \nu}^{(i)} F_{(i)}^{\mu \nu} +\mcl_{rm}
\eeq
where $\mcl_{rm}$ is the Lagrangian for matter and radiation. Here and throughout, the reduced Planck mass is $M_P = 1/\sqrt{8\pi G}$. We assume the fermions associated with the vector fields are not present, having been diluted by inflation or simply too heavy to reach equilibrium with the thermal radiation. As usual, the field strength tensor for the $i^{\rm{th}}$ field is $F_{\mu \nu}^{(i)} = \partial_{\mu} A_{\nu}^{(i)}- \partial_{\nu} A_{\mu}^{(i)}$ and the associated stress-energy tensor is $T_{\mu \nu}^{(i)} = F^{(i)\alpha}_{\mu} F_{\nu \alpha}^{(i)} - g_{\mu\nu} F^{(i)2}/4$. The total stress energy of the system is the sum over the three, $T_{\mu \nu} = \sum_i T_{\mu \nu}^{(i)}$. To proceed, both the metric and the gauge fields are split into a background piece and a linear perturbation: $g_{\mu \nu} = \eta_{\mu \nu} + h_{\mu \nu}$ and $A_{\mu}^{(i)} = \bar{A}_{\mu}^{(i)} + \delta A_{\mu}^{(i)}$.

\subsection{Background}
\label{subsec:background}

We consider a homogeneous, isotropic spacetime with line element $ds^2 = a^2(\tau)(-d\tau^2 + d\vec x^2)$.
The conformal time is $\tau$ and the expansion scale factor $a$ is normalized so that $a_0=1$ today. The background metric evolves according to the Friedmann equation,
\beq
H^2 \equiv \left(\frac{a^\prime}{a^2} \right)^2 = H_0^2 \biggl[\Omega_M \left(\frac{a_0}{a}\right)^{3} + \Omega_R \left(\frac{a_0}{a}\right)^{4} + \Omega_{\Lambda}\biggr]
\label{eqn:friedmann}
\eeq
where $\Omega_M+\Omega_R+\Omega_\Lambda=1$, and $\prime = d/d\tau$ indicates differentiation with respect to conformal time.

In accordance with the symmetries of the RW spacetime, the vector fields must preserve homogeneous and isotropic stress-energy at the background level; the background fields are set up in a ``flavor-space'' locked configuration, 
%
\begin{equation}
  F^{(i)}_{j0} = E_0 \delta^i_j, \qquad F^{(i)}_{k j} = B_0 \epsilon^{ijk} \label{eqn:gaugeflavorspacelock}
\end{equation}
with constants $E_0,B_0$ representing the vevs of the fields. We stress that these are not the electromagnetic fields of the Standard Model. For each flavor of the U(1) there is an ``electric" field and co-parallel ``magnetic" field; each flavor is aligned with one of the principal spatial directions. We note that the two triads must be so aligned, or anti-aligned, in order for the stress-energy to be isotropic. Put another way, if the electric and magnetic fields of a given flavor were not aligned, there would be a non-zero Poynting vector that would break the rotational symmetry. Later, we also consider the case that the electric- and magnetic-like fields are associated with different flavors, requiring two triplets of U(1) fields.

With this {\it ansatz} the energy density in the background fields is computed via $\rho_{\bar{A}} = u^{\mu} u^{\nu} T_{\mu \nu} = (3/2)(B_0^2 + E_0^2)/a^4$ where $u^\mu$ is the four velocity of an observer at rest with respect to the cosmic frame. The pressure in the $i=x,\, y,\,$ or $z$ direction is $p_{\bar{A}, (i)} = e^{\mu}_{(i)} e^{\nu}_{(i)} T_{\mu \nu} = (1/2)(B_0^2 + E_0^2)/a^4$, the same in all directions, where $e^{\nu}_{(i)}$ are a set of mutually orthogonal basis vectors. The equation of state is $w=1/3$, but these stationary fields are not thermal radiation. To parameterize the ratio of the U(1) energy density to Standard Model radiation (SMR) energy density, we introduce $R_B,\, R_E$, 
\begin{equation}
    R_B \equiv \frac{1}{2} \frac{B_0^2}{\Omega_{SMR} M_P^2 H_0^2}, \quad R_E \equiv \frac{1}{2} \frac{E_0^2}{\Omega_{SMR} M_P^2 H_0^2}, 
    \label{eqn:RERBdefn}
\end{equation}
where $\Omega_R = \Omega_{SMR}(1 + R_B + R_E)$.

\subsection{Perturbations}
\label{subsec:pert}

Our focus in this scenario is the interaction of GWs with the vector fields, so we begin by considering transverse, traceless, synchronous metric perturbations,  $g_{\mu\nu} = \eta_{\mu\nu} + h_{\mu\nu}$. It will be sufficient to consider a GW propagating in the $+\hat{z}$ direction,
\begin{align}
     h_{\mu \nu} = a^2 (\tau) 
    \begin{pmatrix}
    0 & 0 &0 &0 \\
    0 & h_+(\tau,z) & h_{\times}(\tau,z) &0 \\
    0 & h_{\times}(\tau,z) & -h_+(\tau,z) &0 \\
    0 & 0 &0 & 0 
    \end{pmatrix},
\end{align}
where we allow both $+$ and $\times$ polarizations.
 
Similarly, we consider linear U(1) perturbations, $\delta A_\mu^{(i)}$, propagating in the $+\hat{z}$ direction. These wavelike perturbations can be expressed in terms of two independent polarizations, whereby
\begin{align}
    \delta A_\mu^{(1)} &= \frac{1}{2}M_P \left(0,w_+(\tau,z), w_{\times}(\tau,z),0 \right) \\ 
    \delta A_\mu^{(2)} &= \frac{1}{2}M_P \left(0,w_{\times}(\tau,z),  -w_+(\tau,z),0 \right) \\
    \delta A_\mu^{(3)} &= \left(0,0,0,0 \right).
\end{align}
The factor of $M_P$ makes $w_{+,\times}$ dimensionless, and the factor of $1/2$ is introduced to make the equations of motion more symmetrical.

The equations of motion are obtained from the perturbed field equations, $G_{\mu \nu}  = M_P^{-2} T_{\mu \nu}$ for GWs, and $\nabla_\mu F_{(i)}^{\mu \nu} = 0$ for the vector field excitations. We find that the $+$ and $\times$ polarizations mix, but by switching to a circularly polarized basis,
\begin{eqnarray}
h_{R,L} &=& \frac{1}{\sqrt{2}}(h_+ \pm i h_\times) \\
w_{R,L} &=& \frac{1}{\sqrt{2}}(w_+ \pm i w_\times), 
\end{eqnarray}
the wave equations for the two polarizations separate. For further convenience, we define $u \equiv a h$ and move to Fourier space. The coupled equations of motion are
\begin{align}
&u_{\sigma}^{\prime \prime}  +  \left( k^2 - \frac{a^{\prime \prime}}{a} +2 \frac{B_0^2-E_0^2}{a^2 M_P^2}\right) u_{\sigma } \cr 
& \qquad = \frac{2}{a M_P} \left[ E_0 w_{\sigma }^{\prime} - \sigma k B_0  w_{\sigma } \right], \label{eqn:eomu} \\
&w_{\sigma }^{\prime \prime} + k^2w_{\sigma } \cr 
& \qquad =\frac{2}{aM_P} \left[ E_0 \left( \frac{a'}{a} u_{\sigma } - u_{\sigma }^{\prime} \right) - \sigma k B_0u_{\sigma } \right] ,
\label{eqn:eomw}
\end{align}
where $\sigma = \pm 1$ denotes the right (left) circular polarization. We ignore the effects of photon and neutrino free-streaming and details of the thermal history of the radiation fluid on the GW background \cite{Watanabe:2006qe}.

The above system of equations displays significant new features affecting the propagation of GWs in the cosmological spacetime. To begin, there now appears an effective mass term in the $u$ equation. This term plays an important role during the radiation-dominated era, when $a''/a$ is negligible, leading to suppression (enhancement) of superhorizon modes when $B_0^2$ is greater (less) than $E_0^2$. Hence, a blue (red) tilt will be induced in a scale-invariant spectrum. Next, the GWs couple to the wave excitations of the U(1) fields. Interconversion between the two, $u$ and $w$, will deplete a spectrum of GWs across a range of frequencies. This will imprint periodic features onto an otherwise power-law spectrum. Last, the equations of motion differ in chirality. In the presence of both $E_0$ and $B_0$ fields, right-circular polarized waves propagate differently than left-circular. Consequently, the processed spectrum will exhibit an excess handedness as a function of frequency. Together, these three features will transform a primordial spectrum and leave interesting targets for observation and experiment.

Finally, we note that scalar perturbations of the flavor-space locked configuration may be equivalently described as perturbations of an $E$- and $B$-fluid, both with $w = \delta p/\delta \rho = 1/3$ and a scalar shear $\sigma$. Details are given in Appendix \ref{AppendixBackgroundStability}. The equations of motion are identical to the color electrodynamics case previously studied \cite{Bielefeld:2015daa}. There, it was shown that the scalar perturbations are stable and leave negligible imprint on the CMB, given the bounds on the relic energy density. Hence, we will not consider the scalar perturbations further in our analysis.

\subsection{Stochastic Gravitational Wave Background}
\label{subsec:sgwb}

The observable of interest that displays these new features is the GW spectrum $\Omega_{GW}$. This quantity is defined as the energy density in GWs, $\rho_{GW}$, per logarithmic frequency interval, $\ln k$, normalized by the critical density, $\rho_c$ \cite{Watanabe:2006qe},
\begin{align}
    \Omega_{GW}(k,\tau) &\equiv \frac{1}{\rho_c}\frac{d \rho_{GW}}{d \ln k} \label{eqn:OmegaGWdefn} \\
    &= \frac{1}{12 (H a)^2} \sum_{\sigma} \Delta^{2}_{h_{\sigma},\rm{prim}} \left|  \mathcal{T}_{h_\sigma,k}^{\prime} (\tau) \right|^2.  \label{eqn:OmegaGW}
\end{align}
In the second equality we introduce the GW transfer function $\mathcal{T}$,
\begin{equation}
    \mathcal{T}_{h_\sigma,k}(\tau) \equiv h_{k,\sigma}(\tau)/h_{k,\sigma}^{\rm{prim}}, 
\end{equation}
which connects the initial GW amplitude to the later, processed amplitude. The primordial GW power spectrum is 
\begin{equation}
    \Delta_{h_{\sigma},\mathrm{prim} }^2(k) \equiv 2 \frac{k^3}{2\pi^2} \left|h_{\sigma,k,\mathrm{prim}} \right|^2
\end{equation} 
for each polarization. In the standard case of slow-roll inflation, $|h_{\sigma,k,{\rm prim}}|^2 = H_I^2/k^3 M_P^2$ where $H_I$ is the Hubble expansion rate during inflation at horizon exit of a mode with comoving wavenumber $k$.

The spectrum of U(1) excitations can be described similarly, 
\begin{align}
\Omega_{\delta A}(k,\tau) &\equiv \frac{1}{\rho_c} \frac{d \rho_{\delta A}}{d \ln k}   \\
 &= \frac{k^3}{12 \pi^2 (H a^2)^2} \sum_{\sigma} \left|  w_{\sigma,k}^{\prime} (\tau) \right|^2  \\
 &= \frac{1}{12 (H a^2)^2} \sum_{\sigma} \Delta^{2}_{h_{\sigma},\rm{prim}} \left|  \mct_{w_\sigma,k}^{\prime} (\tau) \right|^2. \label{eqn:OmegadeltaA}
\end{align}
Here, we cast the U(1) spectrum in terms of the primordial GW power spectrum, and define the $w$ transfer function $\mathcal{T}_{w_\sigma,k}(\tau) \equiv w_{k,\sigma}(\tau)/h_{k,\sigma}^{\rm{prim}}$, also normalized by the primordial GW amplitude.

We note that the calculation of the energy densities invokes spatial averaging over lengths much larger than the wavelength. Although this procedure makes sense for waves that are well within the horizon, we nevertheless apply the above expressions to wavelengths approaching the Hubble radius. Furthermore, at high frequencies we apply a time-averaging filter to obtain the envelope of the spectrum, $\Omega_{GW,\mathrm{env}} \equiv 2 \langle \Omega_{GW} \rangle_{\tau}$, where the time average $\langle \dots \rangle_{\tau}$ is to be taken over a period of oscillation, $\tau = 2\pi/k$. In subsequent figures, we plot the full oscillatory spectrum $\Omega_{GW}$ up to $k\simeq k_{eq}$. At higher wavenumbers we plot only the envelope.
 
\subsection{Initial Conditions and Parameter Choices}
\label{subsec:init}

We prepare the background spacetime in a radiation-dominated universe following reheating. We use cosmological parameters $(H_0,\,\Omega_M,\, \Omega_{SMR}) = \left(67.66\,\mathrm{km}\,\mathrm{s}^{-1}\,\mathrm{Mpc}^{-1},\,0.3111,\,9.138 \times 10^{-5} \right)$ obtained from the Planck 2018 TT, TE, EE, lowE, lensing, BAO best fit values  \cite{Aghanim:2018eyx}. The energy density of the U(1) fields is set by choosing values of $R_B,\,R_E$. This additional, relativistic energy density is constrained by Big Bang Nucleosynthesis (BBN) to add no more than $\Delta N_{\rm eff} < 0.43$~(2$\sigma$) extra massless neutrinos \cite{Fields:2019pfx}. CMB constraints on extra species, which require some assumptions about the clustering properties, yield $\Delta N_{\rm eff} < 0.3$~(95\% CL) \cite{Aghanim:2018eyx}. A joint BBN-CMB analysis gives $\Delta N_{\rm eff} < 0.168$~(95\% CL) \cite{Fields:2019pfx}. 
We will use this tighter bound to set a limit on the energy density. Hence, expressing $R_B,\,R_E$ in terms of the effective number of neutrino species,   
\begin{equation}
    R_B + R_E = \frac{\frac{7}{8}\left(\frac{4}{11}\right)^{4/3}\Delta N_{\rm eff} }{1 + \frac{7}{8}\left(\frac{4}{11}\right)^{4/3}N_{\rm eff} },
\end{equation}
where $N_{\rm eff}=3.046$, we obtain the upper limit $R_B+R_E < R_{max} = 0.0225$~(95\% CL).

We assume the existence of a spectrum of primordial GWs generated by slow roll inflation. We set the spectrum amplitude at the CMB pivot scale, $k_* = 0.05 \, \mathrm{Mpc}^{-1}$, to be consistent with bounds on the tensor-to-scalar ratio $r_*$ and the normalization of the scalar spectrum, $A_S = 2.101^{+0.031}_{-0.034}\times 10^{-9}$ \cite{Aghanim:2018eyx}. A Planck 2018 analysis yields an upper bound $r_* <  0.06$~(95\% CL) \cite{Akrami:2018odb}, whereas a subsequent study places the upper limit at $r_* < 0.044$~(95\% CL) \cite{Tristram:2020wbi}. To be concrete, we will use $r_* = 0.04$ to set the spectrum amplitude, $\Delta^2_{h_\sigma, {\rm prim}}(k_*) = H_I^2(k_*)/M_P^2 \pi^2$, whereby summing over polarizations we obtain $H_I(k_*) \approx 2.0 \times 10^{-5} M_P$. The amplitude at other wavelengths, ranging from the present-day Hubble radius down to the Hubble radius at the end of reheating, depends on the details of both the inflationary and reheating scenarios. For simplicity, in this work, we make the gross assumptions that the primordial spectral tilt is vanishingly small, $n_T \ll 1$, and that reheating is instantaneous. Hence, we initialize the system at the end of reheating, at conformal time $\tau_{RH} = 1/(a_{RH} H_{RH})$, where $a_{RH} = (\Omega_{R} H_0^2/H_{RH}^2)^{1/4}$ and $H_{RH} = H_I$. For simplicity, we ignore the change in number of degrees of freedom of the radiation fluid. For the perturbations $u_{\sigma,k}$ and $w_{\sigma,k}$, we focus our attention on modes that are outside of the horizon at the end of reheating. This sets the maximum frequency that we consider $k_{RH} = a_{RH} H_{RH}$. We initialize the perturbations so that the GW is frozen outside the horizon, as it would be in the absence of the relic vector fields, while the vector field perturbation is frozen at zero. The transfer functions cast in terms of the variables $h$, or $u$, and $w$ have initial conditions
\begin{align}
    (\mathcal{T}_{h_\sigma,k} ,\mathcal{T}_{h_\sigma,k}^\prime,\mathcal{T}_{w_\sigma,k} ,\mathcal{T}_{w_\sigma,k}^\prime)_{RH}=& \, (1,0,0,0)_{RH}, \\
    (\mathcal{T}_{u_\sigma,k} ,\mathcal{T}_{u_\sigma,k}^\prime,\mathcal{T}_{w_\sigma,k} ,\mathcal{T}_{w_\sigma,k}^\prime)_{RH}=& \, (1,a H,0,0)_{RH}. \label{eqn:TuTwICs}
\end{align}
We are now prepared to evolve the system of equations.

\subsection{Numerical Evolution in the High Frequency Regime}
\label{subsec:numerics}

To compute the GW spectral energy density across a broad range of frequencies we must evolve  Eqs.~(\ref{eqn:friedmann}), (\ref{eqn:eomu}) and (\ref{eqn:eomw}) for the transfer functions of $u,\,w$. For each Fourier mode $k$, we evolve the system of equations from the end of reheating, $\tau_{RH}$, to today, $\tau_0$. We sample approximately 27 decades in wavenumber, $k \in [10^{-5},\, 10^{22}]~\mathrm{Mpc}^{-1}$, sampling linearly for $k \lesssim k_{eq}$ and logarithmically for greater values. Each point in the $(k,\Omega_{GW}(k,\tau_0))$ plane thus corresponds to a single mode evolution of the equations of motion. Accurately tracking the oscillatory behavior of every mode once it enters the horizon presents a computational challenge, so some comments on the procedure are in order.

Solving the system of equations exactly must be done numerically. However, the numerical integration becomes computationally unfeasible once the modes are deep within the horizon, $k  \gg \mathcal{H}$, due to the small step size required to resolve the high frequency oscillations of the GW and U(1) modes. To circumvent this issue, we evolve each mode numerically from $\tau_{RH}$ until it is deep within the horizon at some time $\tau_m \sim 10^4/k$. At this point we match the numerical solution to an approximate analytic solution. In this high frequency limit, $k \tau \gg 1$, a WKB approximation is made, setting $u(\tau) = \mathrm{Re}\left[U(\tau)e^{ik(\tau-\tau_m)} \right]$ and assuming several conditions: (1) the mode is deep within the horizon, $k\gg \mch$; (2) the envelope function $U$ varies slowly compared to $k$, $\left|U\pr \right| \ll k \left|U\right|$; (3) the $a\prpr/a$ term is negligible (e.g. in a radiation-dominated background). This analytic solution captures the remaining evolution from $\tau_m$ to $\tau_0$, whereby the envelope function $U(\tau)$ is
\beq
 U(\tau)= U_+ e^{i \omega \int_{\tau_m}^{\tau} d\tau^{\prime}/a(\tau^{\prime}) } + U_- e^{-i \omega \int_{\tau_m}^{\tau} d\tau^{\prime}/a(\tau^{\prime})  } \label{highfreqanalyticUsoln}
\eeq 
and $\omega = \sqrt{({B}_0^2 + {E}_0^2)/M_P^2}$ (with an analogous solution for $w(\tau)$). The full details of this high-frequency analytic solution and matching technique are presented in Appendix~\ref{AppendixHighFreq}. 

\begin{figure}[h!]
    \includegraphics[width=0.95\linewidth]{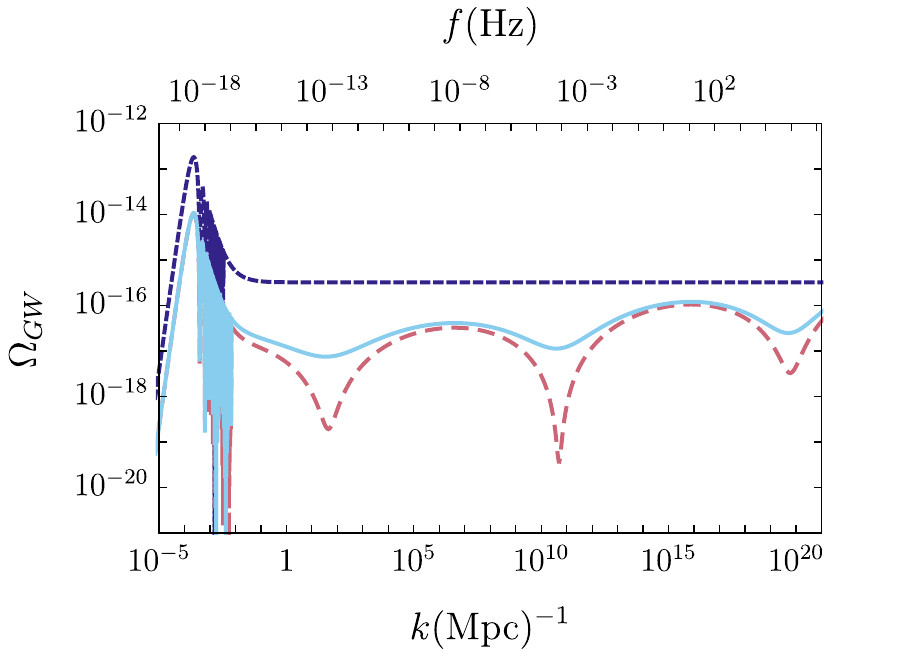}
    \includegraphics[width=0.95\linewidth]{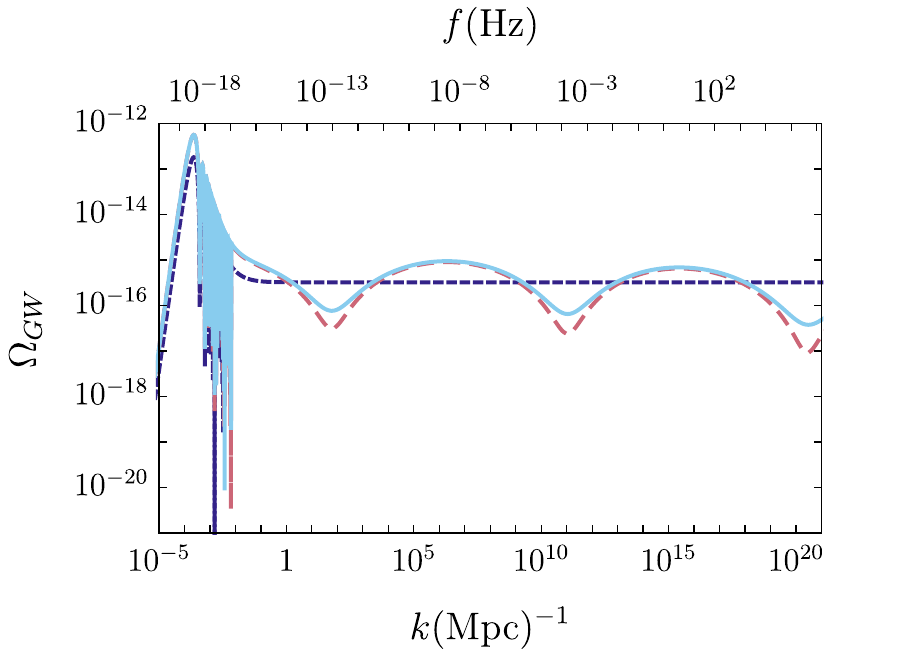}
    \caption{A comparison of $\Omega_{GW}$ between the standard case (dark blue, short dashed) and this model in a) a B-dominant case, upper panel, $(R_B,R_E)=(10^{-2},10^{-3})$; and b) an E-dominant case, lower panel, $(R_B,R_E)=(10^{-3},10^{-2})$. Both polarizations in this model are shown: $\sigma = 1$ (light red, long dashed) and $\sigma = -1$ (light blue, solid).}
\label{fig:EBdom}
\end{figure}

Here we briefly consider the standard case, without the U(1) fields, to illustrate this procedure. The equation of motion for the GW is 
\begin{align}
u''_{\mathrm{std}} + \left(k^2 - \frac{a''}{a}\right)u_{\mathrm{std}} = 0. \label{hstdeom}
\end{align}
The numerical solution requires prohibitively small step sizes once the mode is deep within the horizon, but this can be easily circumvented with the analytic approximations described above. Once a mode is deep inside the horizon, we use the high frequency {\it ansatz} for $u(\tau)$ to find a matching solution to the exact evolution, valid for $\tau > \tau_m$. In this standard case, the envelope function $U$ is a constant. The resulting spectral energy density, shown in Fig.~\ref{fig:EBdom}, is easily characterized. For superhorizon modes $k<a_0 H_0$, $\Omega_{GW} \propto k^4$; for modes that entered the horizon since radiation-matter equality, $a_0 H_0<k<k_{eq}$, $\Omega_{GW,{\rm env}} \propto k^{-2}$; and for modes that entered the horizon prior, $k>k_{eq}$, the envelope is a constant $\Omega_{GW,{\rm env}} \simeq \Omega_R H_I^2/12 \pi^2 M_P^2$. 

In more general cases, the envelope functions for $u,\, w$ satisfy coupled first-order differential equations; the values of $u,\,w$ at the matching point are used to provide boundary conditions for the envelope solution. The analytic solution then captures the evolution from $\tau_m$ to $\tau_0$.

\section{Results}
\label{sec:Results}

We compute the present-day GW spectrum $\Omega_{GW}$ in a variety of cases, to demonstrate the impact of the U(1) gauge fields. We begin with a B-dominant case, $(R_B,R_E) = (10^{-2},10^{-3})$, and an E-dominant case, $(R_B,R_E) = (10^{-3},10^{-2})$, as illustrated in the upper and lower panels respectively of Fig.~\ref{fig:EBdom}. Both circular polarizations are shown. All curves are generated assuming a standard inflationary scenario with the same inflationary scale $H_I \approx 2.0\times 10^{-5} M_P$ and a flat primordial GW spectrum, $n_T = 0$. Upon inspection, it is seen that the GW spectrum acquires three new features compared to the standard case: a tilt, a net circular polarization, and oscillations with logarithmic frequency. Let us now consider each feature in turn.

\begin{figure}[b!]
\vspace{4pt}
    \includegraphics[width=0.45\textwidth]{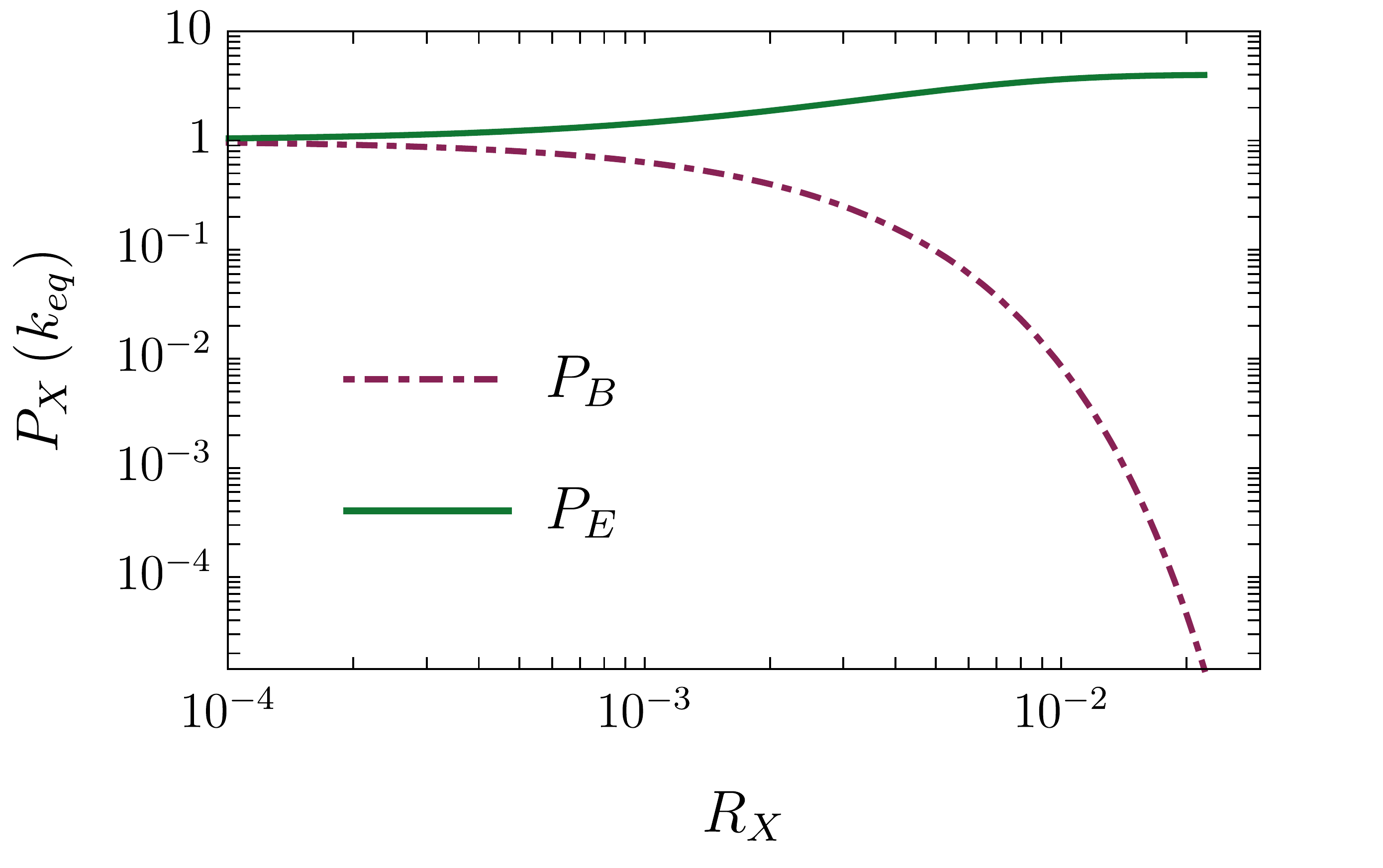}
    \caption{The behavior of the tilt functions $P_B$ (dark red, dot-dashed) and $P_E$ (green, solid) at $k = k_{eq} = k_{piv}$ across a range of background values $R_X$. In the B-dominant case the suppression factor reaches a minimal value of $P_B \approx 10^{-5}$ for $R_B = R_{max}$. The E-dominant case reaches a maximal value of $P_E \approx 4$ for $R_E = R_{max}$.}
    \label{fig:PBPEplot}
\end{figure}

\subsection{Tilt}\label{subsec:tilt}

The tilt of the spectrum is a consequence of the effective mass term contributed by the U(1) background fields, $E_0$ and $B_0$. The mass term, seen in Eq.~(\ref{eqn:eomu}), modifies the dispersion relation during the radiation era, so that low frequency, long wavelength modes are no longer frozen outside the horizon. Rather, superhorizon modes are amplified for $E_0>B_0$, and suppressed for $B_0 > E_0$. Because modes with lower frequencies enter the horizon later, this modified superhorizon evolution has a larger cumulative effect for longer wavelengths. Hence, $E_0 > B_0$ will impart a red tilt to modes $k \gtrsim k_{eq}$, and a blue tilt for $B_0 > E_0$. The magnitude of the enhancement or suppression $P_X$ for $X=B,\,E$ at wavenumber $k_{eq}$ is illustrated in Fig.~\ref{fig:PBPEplot}. The maximum enhancement is a factor of $4$ in the E-dominant case, whereas the B-dominant suppression can push the spectrum down by orders of magnitude. Because this phenomenon produces a constant offset for $k<k_{piv} \approx k_{eq}$, the enhancement or suppression affects wavelengths relevant for the CMB. 

\begin{figure}[t!]
    \includegraphics[width=0.45\textwidth]{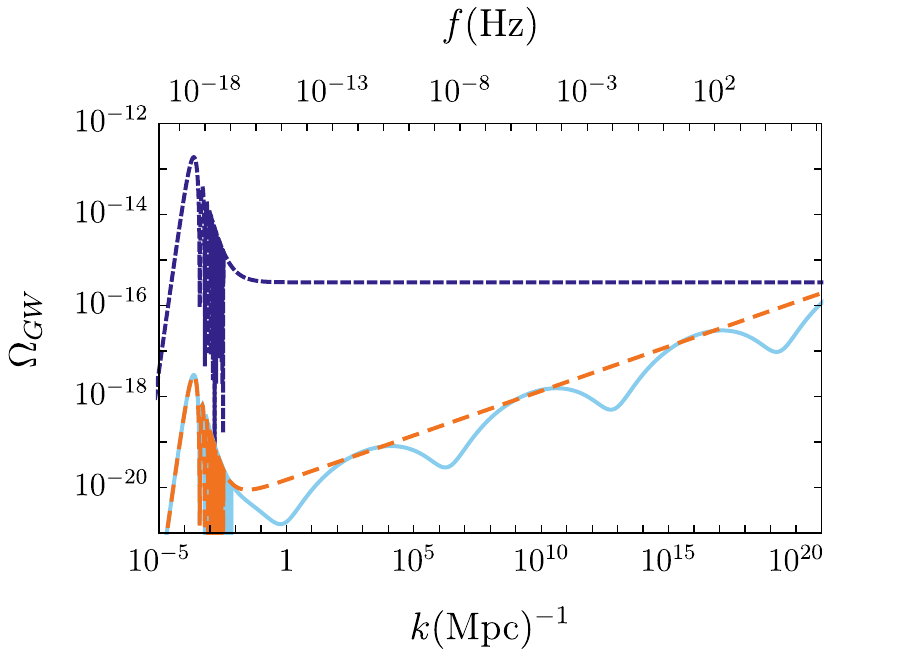}
    \caption{Demonstration of the approximation of the modified GW spectrum using the tilt function, Eqs.~(\ref{tiltspectrumformula}-\ref{PBtiltformula}), in a scenario with $(R_B,R_E)=(R_{max},0)$. The approximation (orange, long dashed) of the spectrum as the product of the standard spectrum, $\Omega_{GW,\rm{std}}$ (dark blue, short dashed), and a tilt function, $P_B$, reliably predicts the tilt of the numerical result (light blue, solid).}
    \label{fig:PBtiltfit}
\end{figure}

The degree of tilt can be predicted by solving the equations of motion in the limit $k \rightarrow 0$ and taking the ratio of the transfer function amplitude at horizon entry $k\tau_{entry} =1$ in the modified and standard cases, $P=\left| \mct_u (\tau_{entry}) / \mct_{u,\rm{std}} (\tau_{entry}) \right|^2$. This ratio $P$ describes the frequency-dependent modification of the standard spectrum produced by the gauge fields. Here we quote the result for the simple case where one of the background fields is zero, since the tilt is only significant when one background field dominates over the other. In these cases, momentarily setting aside other effects, the newly tilted spectrum can be well-approximated as a product of the standard spectrum and a tilt function 
\begin{equation}
    \Omega_{GW,X} \simeq \Omega_{GW,\mathrm{std}} P_X(k).
    \label{tiltspectrumformula}
\end{equation}
The tilt functions in this case are given by piecewise continuous functions
\begin{align}
P_B (k)  &={\cal A}_R^2\lp \frac{ {\rm max}(k,k_{piv}) }{k_{RH}} \rp^{ n_B} \label{PBtiltformula}  \\
P_E (k)  &= 
       \lb 2+ {\cal A}_R\lp \frac{ {\rm max}(k,k_{piv}) }{k_{RH}} \rp^{\frac{1}{2}  n_E } \rb^2,
       \label{PEtiltformula} 
\end{align}
where $k_{piv}$ is a pivot scale, $n_B = 1- \sqrt{1-16 R_B}$, $n_E = 1- \sqrt{1-16 R_E}$, $k_{RH} = a_{RH} H_{RH}$, and
\begin{equation}
    {\cal A}_R = \frac{R_B - R_E}{R_B + R_E}  \frac{1+\sqrt{1-16(R_B+R_E)}}{2\sqrt{1-16(R_B+R_E)}}.
\end{equation}
See Appendix \ref{AppendixSuperhorizon} for details of this calculation.
The modification to the standard case is a power law for $k_{piv}<k<k_{RH}$, while for $k<k_{piv}$ it becomes a constant offset. Intuitively $k_{piv}$ should be close to $k_{eq}$, because after radiation-matter equality, the U(1) fields dilute rapidly compared to the background and no longer affect the superhorizon evolution of the GWs. Hence all modes that enter the horizon after equality have the same modification to their amplitude.  A comparison of the above model to the exact spectrum is shown in Fig.~\ref{fig:PBtiltfit}. We find good agreement between the approximate tilt functions and the numerical results when using $k_{piv} = k_{eq}$.

\subsection{Chirality}\label{subsec:chirality}

Starting from a GW spectrum with equal amplitude left- and right-circular polarizations, the presence of the gauge field vevs in this scenario will cause a net chirality to develop over time. This effect requires the presence of both background gauge fields $E_0$ and $B_0$, and is relevant as the wave enters the horizon. The necessity of $B_0$ is an obvious consequence of the axial nature of the magnetic field. However, $E_0$ is also necessary, if only to provide a reference against which left and right can be defined. This can be seen by recognizing that when $E_0 = 0$, the equations of motion for the two polarizations differ only by a minus sign, $u_{R} = - u_{L}$, which has no consequence. Furthermore, in both the $k \rightarrow 0$ and $k \gg {\cal H}$ limits, for wavenumbers that are far from the effective Hubble scale introduced by the gauge fields, the polarization $\sigma$ drops from the dynamical equations. 

\begin{figure}[b!]
    \includegraphics[width=0.45\textwidth]{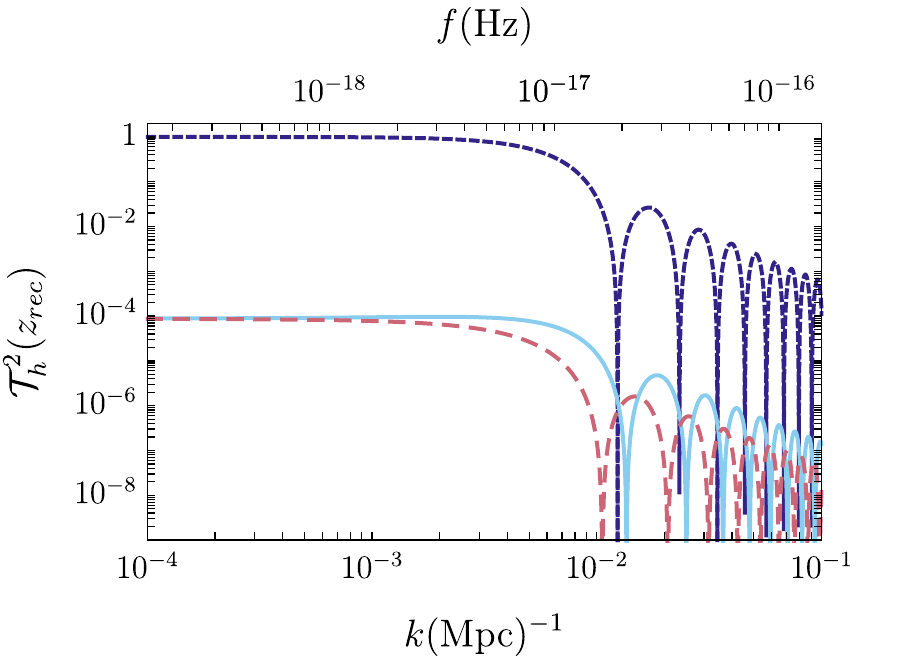}
    \caption{The squared GW transfer function $\mct_h^2$ at recombination with $(R_B,R_E)/0.02 = (1-10^{-3},10^{-3})$ for both the $\sigma = -1$ (light blue, solid) and $\sigma=1$ (light red, long dashed) polarizations, compared to the standard case (dark blue, short dashed).}
    \label{fig:GWtrasnferfunc}
\end{figure}
 
Interestingly, considering the cases in Fig.~\ref{fig:EBdom}, the maximum chirality in the B-dominant case can be quite significant, with the amplitude of the two polarizations differing by over an order of magnitude at some frequencies, whereas the chirality is weaker when the dominance is reversed; the chirality depends on the comparison of $B_0$ relative to $E_0$. To further characterize the chirality in this model, in Fig.~\ref{fig:GWtrasnferfunc} we compare the GW transfer function at CMB formation in a B-dominant example to the standard case. The suppressed GW amplitude and the net circular polarization are both consequences of the interactions between the U(1) gauge fields and the primordial GWs. The $\sigma=\pm 1$ polarization enters the horizon later (earlier) and is relatively suppressed (amplified). In this scenario the resultant B mode spectrum of the CMB would have a net circular polarization.

\subsection{Oscillations}\label{subsec:oscillations}

\begin{figure}[t!]
    \includegraphics[width=0.45\textwidth]{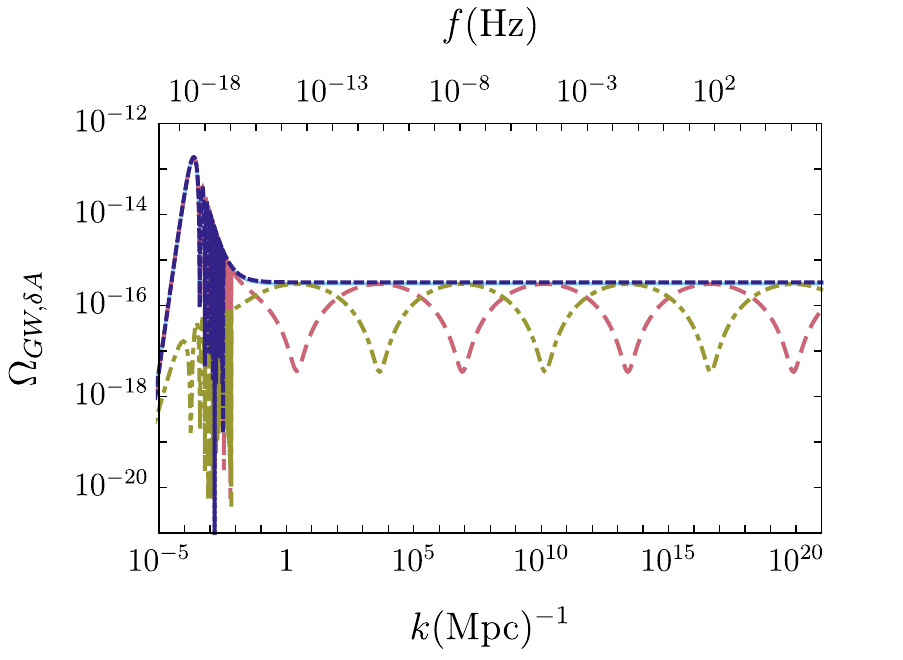}
    \caption{The spectral energy densities of the GW (light red, long dashed) and U(1) perturbations (dark yellow, dot-dashed) for the right-handed ($\sigma = 1$) polarizations in an $R_B = R_E = R_{max}/2$ case. For comparison we also show $\Omega_{GW,\rm{std}}$ (dark blue, short dashed) and the sum of the GW and U(1) spectral energy densities (light blue, solid); note these two curves are nearly on top of one another.}
    \label{fig:gwosc}
\end{figure}

The oscillation of the spectral energy density over broad frequency ranges is best understood by considering the spectra of both GWs and the U(1) perturbations $\delta A$ together. Fig.~\ref{fig:gwosc} makes clear that the oscillations in $\Omega_{GW}$ are complementary to those in $\Omega_{\delta A}$: as one rises, the other falls, and the sum yields a flat spectrum. This feature is a consequence of the interconversion between the GWs and the U(1) perturbations in the presence of the background $E_0$ and $B_0$ fields. The phenomenon is referred to alternately as the Gertsenshteyn effect \cite{Gertsenshteyn1962}, photon-graviton conversion \cite{Poznanin1969,Boccaletti1970,Zeldovich1974}, and more generally as GW - gauge field oscillations \cite{Caldwell:2016sut}. We now discuss the main idea underpinning this effect.

The interconversion is seen in the amplitude of the GW and U(1) perturbations during the radiation era. Subhorizon, the modes oscillate rapidly  with frequency $k$, modulated by the more slowly varying envelope 
\begin{equation}
    e^{i\omega \int^{\tau}_{\tau_m}d\tau\pr/a(\tau\pr)}.
\end{equation}
The lower bound of integration is set by horizon entry $\tau \sim k^{-1}$. The upper bound is set by radiation-matter equality, when the interconversion process effectively turns off. Hence, modes that enter the horizon earlier (later) will accumulate more (less) phase modulation. Because the scale factor evolves as $a \propto \tau$ in the radiation era, the phase acquires a logarithmic dependence on wavenumber. This slow modulation survives the time-averaging in the evaluation of the spectral energy density, and imparts the log-scale modulation seen in Fig.~\ref{fig:gwosc}. A detailed derivation is given in Appendix \ref{AppendixHighFreq}.  
 
\begin{figure}[b!]
    \includegraphics[width=0.45\textwidth]{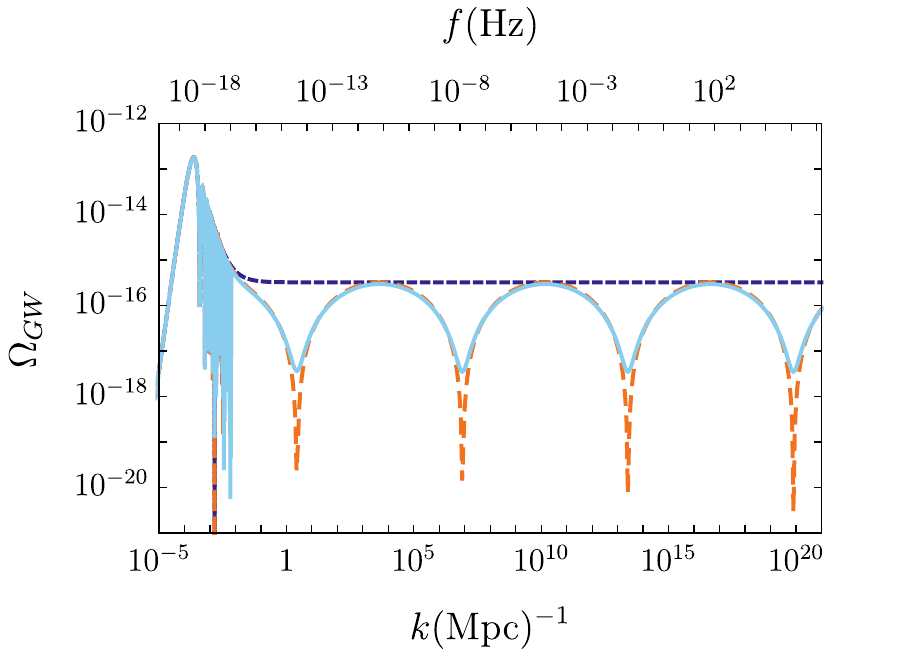}
    \caption{Demonstration of the approximation of the oscillatory GW spectrum using the fitting function, Eq.~(\ref{eqn:oscfitformula}), in a scenario with $R_B=R_E=R_{max}/2$. Using a pivot scale $k_{piv}=k_{eq}/6$, the approximation (orange, long dashed) of the spectrum as the product $\cos^2[\Phi]\times\Omega_{GW,\mathrm{std}}$  reliably predicts the shape of the numerical result (light blue, solid). The standard case (dark blue, short dashed) is included for comparison.}
    \label{fig:gwCosineFit}
\end{figure}

\begin{figure}[t!]
    \includegraphics[width=0.45\textwidth]{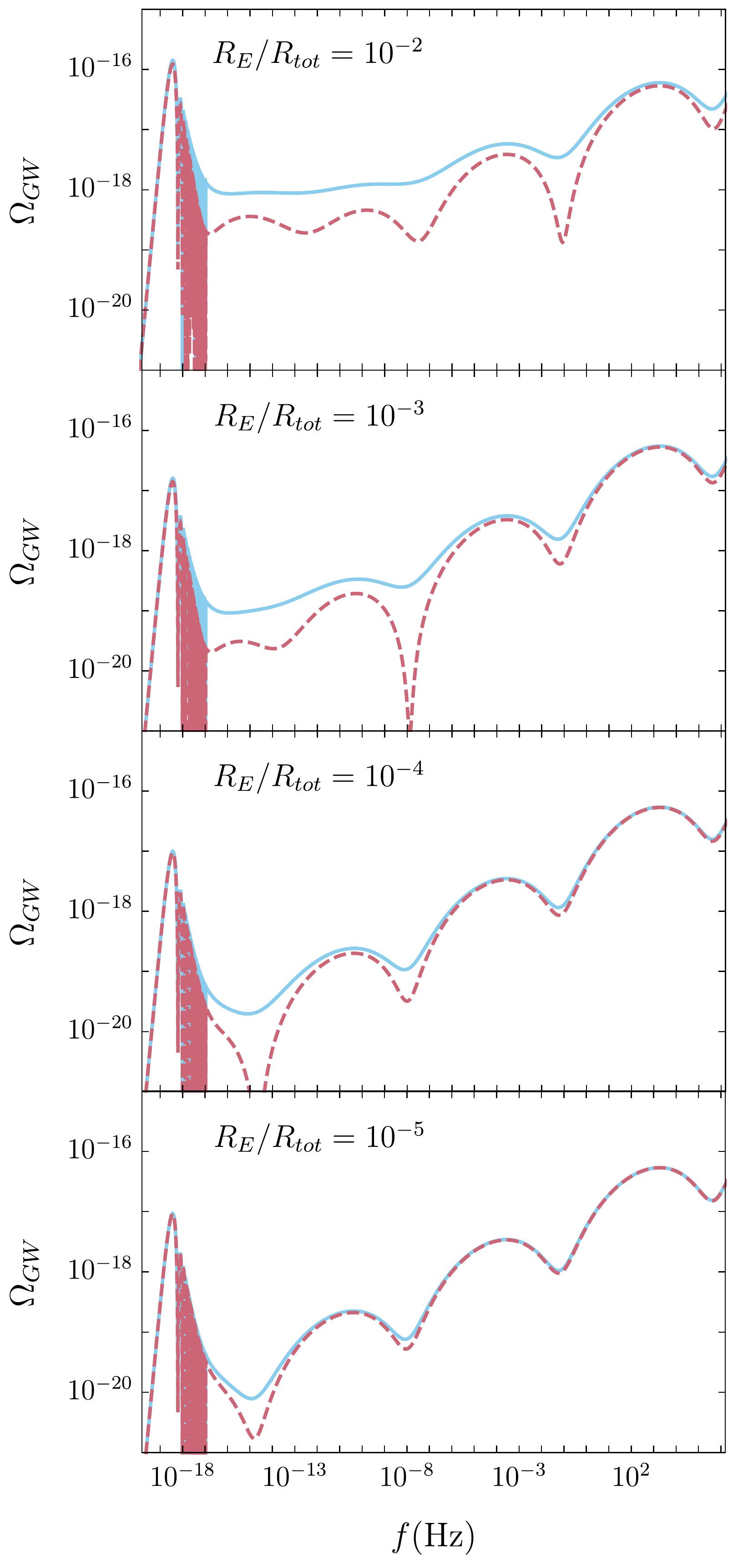}
    \caption{The behavior of the $\sigma = -1$ (light blue, solid) and $\sigma=1$ (light red, long dashed) polarizations of the GW spectra in increasingly B-dominant cases for $R_{tot} = 0.02$.}
    \label{fig:GWpolconverge}
\end{figure}

This analysis allows one to predict the oscillatory shape of the spectra by relating the background U(1) energy density in the model to the frequency dependence. The new oscillatory contribution to $\Omega_{GW}$ can be approximated simplistically by multiplying the  $k \gtrsim k_{eq}$ part of the otherwise power-law spectrum by $\cos^2\Phi$ where
\begin{equation}
    \Phi = \sqrt{2(R_E+R_B)}\ln \left({k}/{k_{piv}} \right), \label{eqn:oscfitformula}
\end{equation}
and $k_{piv}$ is a new pivot scale that should also be close to $k_{eq}$ for the same reasons as in Section \ref{subsec:tilt}. Because of the simplistic nature of this approximation, the pivot scale $k_{piv}$ varies slightly with $R_E,R_B$. When $R_E \approx R_B$, using $k_{piv}/k_{eq} = 1/6$ in Eq.~(\ref{eqn:oscfitformula}) provides an excellent description of the oscillations. In the limiting case where $R_B \rightarrow 0$, we find good results using the empirical relation $k_{piv}/k_{eq} = a (R_E/b)^c$ for $(a,b,c) = (1/5,1/50,-9/50)$. When $R_E \rightarrow 0$, $k_{piv}/k_{eq} = a (R_B/b)^c$ for $(a,b,c) = (1/18,1/50,1/20)$ gives good results. Setting the phase $\Phi$ to an even (odd) multiple of $\pi/2$ gives the frequency for a peak (dip). Adjacent extrema at $k_1$ and $k_2$ are thus related by $\ln(k_1/k_2) = (\pi/2)/\sqrt{2(R_E+R_B)}$. Fig.~\ref{fig:gwCosineFit} shows a comparison of this simple prediction with the detailed numerical calculation.

The simple cosine model of Eq.~(\ref{eqn:oscfitformula}) is effective at capturing the locations of the oscillatory features, but the full structure of the GW spectrum can be far richer in the presence of chirality. We provide an example in the four panels of Fig.~\ref{fig:GWpolconverge}, in which we fix $R_{tot} = R_B +R_E = 0.02$ and progressively lower the fraction of $R_E/R_{tot}$. We observe that the frequency dependence of the GW spectra is complicated, and the spectra are substantially different for the two polarizations. The $\sigma=1$ polarization spectrum is generally lower in amplitude than for $\sigma=-1$; the underlying reason is that for a given k-mode, the $\sigma=1$ transfer function begins to oscillate sooner as it approaches horizon entry. Furthermore, the $\sigma=1$ spectrum appears to have one deep minimum, like a beat frequency, that progressively shifts to lower frequency minima as $R_E/R_{tot}$ is lowered. The behaviour displayed in the figure clearly shows the limitations of the $\cos^2(\Phi)$ model when significant chirality is present. Finally, it is only once $R_E/R_{tot}$ is lowered below $10^{-4}$ that the two polarization spectra begin to converge. Roughly speaking, for larger (smaller) $R_{tot}$, the trend towards convergence occurs at smaller (larger) values of $R_E/R_{tot}$. 

\subsection{Sextet Model}\label{section:sextetmodelresults}
We extend our analysis by considering a scenario in which the electric- and magnetic-like fields originate from distinct gauge groups. Specifically we consider a model variation in which the electric and magnetic backgrounds are distributed among two separate U(1) triplets. This model, which we refer to as a sextet, is very similar to the triplet model, so we keep our discussion brief. Calculation details are presented in Appendix \ref{AppendixSixU1}. 

\begin{figure}[b!]
    \includegraphics[width=0.45\textwidth]{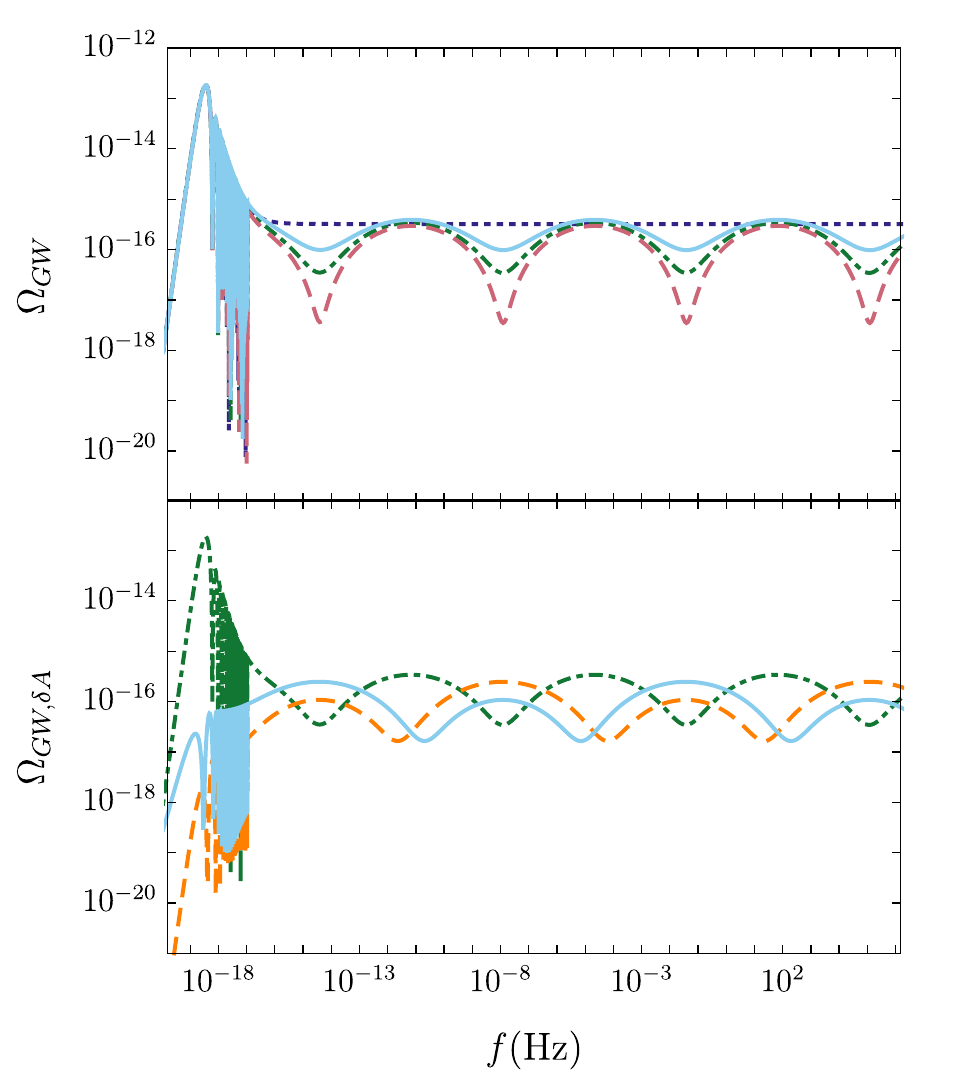}
    \caption{Upper panel: The GW spectrum for a single polarization of the sextet model (green, dot-dashed) compared to that of the $\sigma = -1$ (light blue, solid) and $\sigma=1$ (light red, long dashed)  polarizations in the triplet model, for $R_E = R_B = R_{max}/2 $. The standard case (dark blue, short dashed) is included for comparison. Lower panel: Comparison of spectra in the sextet model for a single polarization of GW (green, dot-dashed), $w$ excitation (orange, long dashed), and $y$ excitation (light blue, solid), for $R_E = R_B = R_{max}/2 $.}
    \label{fig:GW6U1polandosc_f}
\end{figure}

The sextet and triplet models have the same equations of motion in the $k\rightarrow 0$ limit, so the superhorizon behavior, and therefore the tilt in the GW spectrum, is unchanged. The models are also the same in the limit that either background, $E_0$ or $B_0$, goes to zero. However, the sextet model exhibits no handedness, so the excess circular polarization in the transfer function in Fig.~\ref{fig:GWtrasnferfunc}, and the rich features seen in the upper panels of Fig.~\ref{fig:GWpolconverge}, are absent. Rather, the unpolarized GW spectral energy density in the sextet model generally behaves like an intermediate of the enhanced and suppressed polarizations seen in the triplet model. We demonstrate this in the upper panel of Fig.~\ref{fig:GW6U1polandosc_f}.
Furthermore, the sextet model includes two independent wavelike gauge field excitations that mix with the GWs, so the structure of the gauge field spectra can be more complicated. Importantly, the oscillations in the GW spectrum envelope in the two models are described equally well by the approximation Eq.~(\ref{eqn:oscfitformula}), excepting strongly chiral cases in the triplet model. As seen in the lower panel of Fig.~\ref{fig:GW6U1polandosc_f}, the oscillations in the two gauge field excitations are together complementary to the oscillations in the GW spectrum.
\section{Discussion \label{discussion}}

In the present work we have investigated a toy model in which a homogeneous, isotropic configuration
of relic U(1) vector fields interact with primordial GWs throughout the radiation-dominated portion of the post-inflationary Universe. For simplicity, and to isolate the effects of the relics after inflation, we have assumed instantaneous reheating and that the primordial GW spectrum is generated from a standard inflationary scenario, with $n_T =0$. However, it is straightforward to superimpose the effects of the relics on any other primordial spectrum. 

We have focused our study on the imprint left on the GW spectral energy density, $\Omega_{GW}$, finding that the relic fields impart to the spectrum three distinct features: a tilt, a net circular polarization, and oscillations across the high frequency ($k\gtrsim k_{eq}$) spectrum.

\begin{figure}[t!]
    \includegraphics[width=0.45\textwidth]{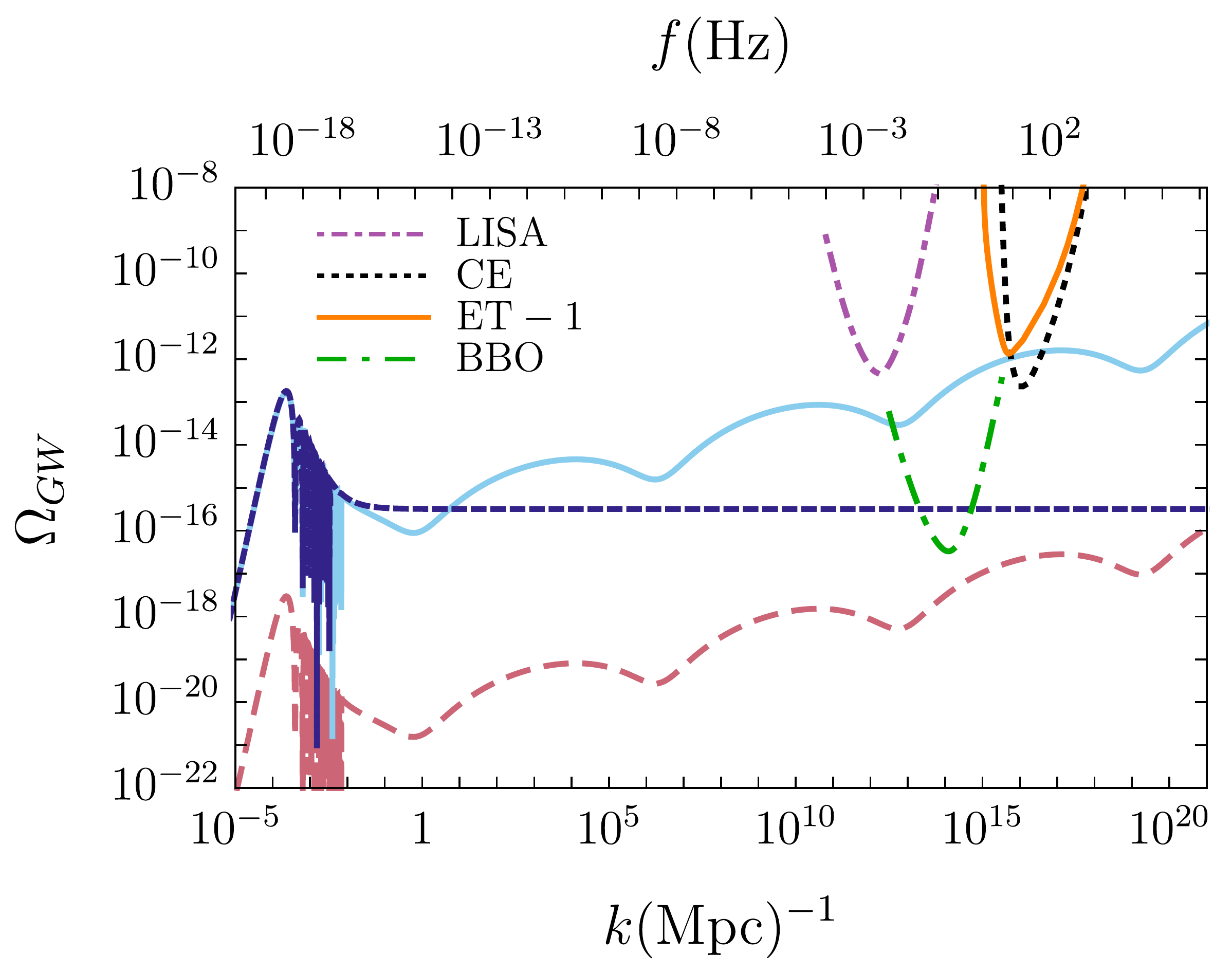}
    \caption{GW spectral density versus wavenumber and frequency are shown, relative to the power-law sensitivity curves for several proposed or futuristic GW observatories. The GW spectra correspond to a B-dominant case, $(R_B,R_E) = (R_{max},0)$, with $H_I = 2\times 10^{-5} M_P$ (light red, long dashed) and $H_I = 2 M_P$ (light blue, solid). This scenario represents the maximal modification to the standard case (dark blue, short dashed) in which $(R_E,R_B)=(0,0)$, given the energy density bounds on the relic cosmic vector fields. Any power-law spectral density that crosses a sensitivity curve is detectable at $\mathrm{SNR}=5$ in a $t_{obs}=4\,\mathrm{yr}$ experiment.}
    \label{fig:Bsuppresed}
\end{figure}

In particular, the tilt, induced by the U(1) background  and demonstrated in Fig.~\ref{fig:Bsuppresed}, is our primary result. Compared to previous work, e.g. \cite{Adshead:2012kp,Maleknejad:2011jw,Bielefeld:2015daa}, which explored a gauge vev that produces a background ``electric" field, the setup presented here contains a new ingredient at the background level in the form of an independent axial ``magnetic" field. This new ingredient allows the GW effective mass term to be positive, altering the dispersion relation and producing the tilt described in Section \ref{subsec:tilt}. In the maximal scenario, the $R_B \gg R_E$ case permits a much stronger tilt than in the reverse case, and so the suppression when $R_B$ dominates can be quite significant. The consequences of such a scenario are noteworthy.   

Broadly speaking this scenario is one of several in which the additional ingredients complicate the relationship between the present day tensor-to-scalar ratio $r$ and the inflationary scale $H_I$. The novel element here is that this effect is a consequence of dynamics after inflation, rather than during it. We have demonstrated that this tilt can reduce the amplitude of $\Omega_{GW}$ by up to $\sim 5$ orders of magnitude at CMB frequencies $k \lesssim k_{eq}$, all while keeping $H_I$ fixed. This is important for inflation model building because it implies that models that predict $r$ values previously considered too large may instead be within observational bounds due to this post-inflationary effect. 

One illustrative example is single field slow-roll inflation with a quadratic potential. Another example is chromonatural inflation \cite{Adshead:2012kp}. Both of these models overproduce GWs for acceptable values of the scalar spectral index, $n_s$. For the free massive field, $r\sim 0.133$. However, $R_B \gtrsim 10^{-1}R_{max}$ would be sufficient to drop the amplitude at CMB scales down below $r \lesssim 0.04$. For the case of chromonatural inflation, the tensor-to-scalar ratio is estimated to be $r \sim {\cal O}(10^2)$. In this case, a larger suppression factor with $R_B \sim 3R_{max}/4$   could bring the model within the $1\sigma$ contour in the $n_s-r$ plane \cite{Ade:2018gkx,Akrami:2018odb}. Fig.~\ref{fig:PBPEplot} and Eq.~(\ref{PBtiltformula}) show the range of suppression possible in this model. In numbers, the range $R_B \in \left[5\times10^{-3}, 10^{-2} \right]$ roughly corresponds to a suppression factor of $P_B \in \left[10^{-1}, 10^{-2} \right]$, which may be enough to revitalize a variety of models with otherwise too strong GW backgrounds. These arguments also apply to models of inflation with axion spectator fields.

The post-inflationary effect of the vector fields on $\Omega_{GW}$ has implications for CMB probes of B-modes as well as direct detection by future GW observatories. Future CMB experiments such as the Simons Observatory \cite{Ade:2018sbj}, LiteBIRD \cite{Hazumi:2019lys}, and CMB-S4 \cite{Abazajian:2016yjj}, with sensitivities approaching $\delta r\sim 10^{-3}$ can constrain the $R_E-R_B$ parameter space in multiple ways. A bound on, or detection of, primordial B-modes may be interpreted  as a joint bound on a model of inflation and the presence of post-inflationary vector fields. Forthcoming CMB experiments are also expected to improve measurement of $\Delta N_\mathrm{eff}$. The GW chirality catalyzed by the vector fields, however, may prove too weak to be detectable with the CMB \cite{Gluscevic:2010vv,Gerbino:2016mqb,Thorne:2017jft}. 

Future GW detectors may find a new target within these scenarios. For illustrative purposes only, in Fig.~\ref{fig:Bsuppresed} we show the power-law sensitivity curves \cite{Thrane:2013oya,Smith:2019wny,Schmitz:2020syl} for the Laser Interferometer Space Antenna (LISA) \cite{Audley:2017drz}, Cosmic Explorer (CE) \cite{Evans:2016mbw}, Einstein Telescope (ET) \cite{Hild:2010id,Punturo:2021ETsensitivity} (assuming a single Michelson interferometer), and the futuristic Big Bang Observer (BBO) \cite{Crowder:2005nr,Harry:2006fi}. Under optimal conditions, one of these detectors could be sensitive to the tilt, chirality, and oscillatory features imprinted in the stochastic GW background.

\vspace{0.5cm}
\acknowledgments

This work is supported in part by U.S. Department of Energy Award No. DE-SC0010386. 
\vfill

\begin{widetext}
\noindent
\appendix 
\section{Scalar Perturbations}
\label{AppendixBackgroundStability}

In this appendix we compute the scalar perturbations in the U(1) vector field model in the background flavor-space locked configuration, Eq.~(\ref{eqn:gaugeflavorspacelock}). For this analysis both the metric and U(1) fields are  perturbed with only scalar degrees of freedom, as linear perturbations in a scalar-vector-tensor decomposition ensure the scalar, vector, and tensor modes do not mix. The U(1) freedom in this model allows us to perform a coordinate rotation such that the Fourier vector points along the $+z$ axis, so the perturbations will be functions of $(\tau,z)$ only. The metric in a flat, RW Universe is then split into a background metric and linear perturbation, $ g_{\mu \nu} = \eta_{\mu \nu} + \delta g_{\mu \nu}$. In a gauge with smooth spatial sections, the flat-slicing gauge, the background metric and perturbations,
\begin{align}
    \eta_{00} &= -a^2(\tau) \\
    \eta_{ij} &= a^2(\tau) \delta_{ij}\\
    \delta g_{00} &= 2 a^2(\tau) \Phi_S(\tau,z) \\
    \delta g_{0i} &=a^2(\tau) \partial_z b(\tau,z),
\end{align}
give the corresponding line element $ds^2 = g_{\mu\nu} dx^\mu dx^\nu$ as
\beq
ds^2 = a^2(\tau) \left[ -(1-2\Phi_S)d\tau^2 -2\partial_z b \, dz d\tau+ \delta_{ij} \, dx^i dx^j\right] .
\eeq
Similarly the U(1) fields are linearly perturbed,  $A^{(i)}_{\mu} = \bar{A}^{(i)}_{\mu} + \delta A^{(i)}_{\mu}$. 
In analogy with the SU(2) case \cite{Maleknejad:2011sq} the U(1) triplet perturbations can be written
\begin{align}
    \delta A^{(i)}_0 &= \delta^{(i)k}\partial_k \dot{Y} + \delta^{(i)j}u_j\\
    \delta A^{(i)}_j &= \delta^{(i)}_{\phantom{j}j} Q + \delta^{(i)k}\partial_{jk}M + \varepsilon^{(i)k}_{\phantom{j}j}\partial_k P + \delta^{(i)k}\partial_j v_k  + \varepsilon^{(i)k}_{\phantom{j}j} w_k +  \delta^{(i)k}t_{jk}.
\end{align}
We consider only scalar perturbations, 
\begin{align}
    \delta A^{(i)}_0 &= \delta^{(i)k}\partial_k \dot{Y}\\
    \delta A^{(i)}_j &= \delta^{(i)}_{\phantom{j}j} Q   + \delta^{(i)k}\partial_{jk}M + \varepsilon^{(i)k}_{\phantom{j}j}\partial_k P,
\end{align}
where the overdot denotes differentiation with respect to cosmic time $t$. Imposing the $(\tau,z)$-only dependence of the perturbations and defining $(W,\delta G, \delta m, \delta C) \equiv (\dot{Y},Q,-\partial_z^2 M,P)$ gives
\begin{align}
    \delta A^{(1)}_{\mu} &= \left( 0 ,\delta G(\tau,z) , \partial_z \delta C(\tau,z) , 0 \right) \\
    \delta A^{(2)}_{\mu} &= \left( 0, - \partial_z \delta C(\tau,z) , \delta G(\tau,z) , 0 \right) \\
    \delta A^{(3)}_{\mu} &= \left( \partial_z W(\tau,z) , 0 , 0 , \delta G(\tau,z) - \delta m(\tau,z)  \right)
\end{align}
which includes all possible independent scalar perturbations to the U(1) fields.
Stress-energy conservation, $\nabla_{\mu} T^{\mu}_{\phantom{a}\nu} = 0$, and the free U(1) equations of motion, $\nabla_{\mu} F^{\mu \nu, (i)} = 0$, give three equations of motion (second order) and one constraint (first order),
\begin{align}
\delta G \prpr + k^2 \delta G - k^2 E_0 b -E_0 \Phi_S \pr  &=0 \\
\delta C \prpr + k^2 \delta C +B_0 \Phi_S -B_0 b\pr &=0 \\
\delta m \prpr +k^2 \delta G - k^2 E_0 b - k^2 W\pr &=0\\
\delta m \pr - \delta G \pr +E_0 \Phi_S -k^2 W &=0.
\end{align}
With these equations we can obtain the fluid variables through the perturbed stress energy tensor \cite{Ma:1995ey}, giving
\begin{align}
    \bar{\rho} \sigma &= \frac{k^2}{a^4}(-E_0)W + \frac{E_0}{a^4}\delta m\pr + \frac{k^2}{a^4}B_0 \delta C \\
    (\bar{\rho} + \bar{P})\theta &= 2 \frac{k^2}{a^4}(-E_0)\delta G \pr +2 \frac{k^2}{a^4}B_0 (\delta C \pr -B_0 b)\\
    \delta \rho &= \frac{-E_0}{a^4}(3\delta G\pr - \delta m\pr) - \frac{k^2}{a^4} E_0 W  + 3\frac{E_0^2}{a^4} \Phi_S - 2\frac{k^2}{a^4}B_0 \delta C
\end{align}
where the energy density and pressure at background and linear order satisfy the equation of state of radiation, $\bar{P}/\bar{\rho} = 1/3 = \delta P/\delta \rho$. The evolution of the fluid variables is given by 
\begin{align}
    \delta \pr &= -\frac{4}{3}\left( \theta +k^2 b \right)\\
    \theta \pr &= k^2 \left( \frac{1}{4}\delta -\Phi_S -\sigma \right)
\end{align}
where $\delta \equiv \delta \rho /\bar{\rho}$.

Now we transform to the conformal Newtonian (cN) gauge and express the fluid equations in that gauge. The metric in the cN gauge is
\begin{align}
    g_{00} &= -a^2 (1+2\psi_{cN}) \\
    g_{0i} &= 0 \\
    g_{ij} &= a^2(1-2\phi_{cN})\delta_{ij}.
\end{align}
The cN gauge (with coordinates $\hat{x}^{\mu}$) and our gauge (with coordinates $x^\mu$) are related by a simple coordinate change $x^0 = \hat{x}^0 + b$, with no transformation of the spatial coordinates. This yields the relationships \cite{Ma:1995ey} between the metric perturbations in the two gauges, 
\begin{align}
     \mch b &= -\phi_{cN} \label{eqn:bphiCN}\\
    -\Phi_S + \mch b + b\pr &= \psi_{cN} \label{eqn:PhiGpsiCN}.
\end{align}
Under coordinate transformation, the stress energy tensor and thus the fluid variables also transform. It is simplest to proceed by calculating the gauge-invariant (gi) perturbations, which are equivalent to those in the cN gauge \cite{Mukhanov:1990me}. With this we can compute the gauge invariant (equivalently, cN) fluid variables,
\begin{align}
    ^{(gi)} \delta  &= \delta_{cN}  =\frac{3E_0^2 \Phi_S - k^2 E_0 W -E_0(3\delta G\pr - \delta m\pr)  - 2B_0 k^2 \delta C}{3(E_0^2 +B_0^2)/2} + 4\phi_{cN} \\
    ^{(gi)}\theta &= \theta_{cN}=  \frac{-k^2 E_0 \delta G +k^2 E_0 b +k^2 B_0 \delta C\pr}{E_0^2 +B_0^2} \\ 
    ^{(gi)} \sigma &= \sigma_{cN}=\frac{-k^2 E_0 W + E_0\delta m\pr + k^2 B_0 \delta C}{3(E_0^2 +B_0^2)/2},
\end{align}
where one can replace and $b$ and $\Phi_S$ with cN variables using Eqs.~(\ref{eqn:bphiCN}) and (\ref{eqn:PhiGpsiCN}). These cN fluid variables obey
 \begin{align}
     ^{(gi)}\delta\pr &= \delta_{cN}\pr  = -\frac{4}{3}\theta_{cN}  + 4\phi_{cN} \pr \\
     ^{(gi)}\theta\pr &= \theta_{cN}\pr  = k^2 \frac{1}{4} \delta_{cN} - k^2 \sigma_{cN} + k^2 \psi_{cN}\\
     ^{(gi)}\sigma\pr &= \sigma_{cN}\pr = \frac{2}{3}\theta_{cN}.
 \end{align}
 These are analogous to the first two equations in (63) of Ref.~\cite{Ma:1995ey} with a modified shear component, in accordance with what was found in the color electrodynamics case \cite{Bielefeld:2015daa}.

\section{High Frequency GW-U(1) Solutions and Computation of $\Omega_{GW}$}
\label{AppendixHighFreq}
In this section we develop an approximate analytic solution to the equations of motion~(\ref{eqn:eomu}-\ref{eqn:eomw}) in a high frequency limit. This solution is used to compute $\Omega_{GW}$ given the computational challenges described in Section \ref{subsec:numerics}. This solution also demonstrates the origins of the GW envelope, Eq.~(\ref{highfreqanalyticUsoln}), and the consequent oscillations in $\Omega_{GW}$ discussed in Section \ref{subsec:oscillations}. The equations of motion are
\begin{align}
&u_{\sigma}^{\prime \prime}  +  \left( k^2 - \frac{a^{\prime \prime}}{a} +2 \frac{(B_0^2-E_0^2)}{a^2 M_P^2 }\right) u_{\sigma } = \frac{2}{a M_P} \left[ E_0 w_{\sigma }^{\prime} - \sigma k B_0  w_{\sigma } \right] \\
&w_{\sigma }^{\prime \prime} + k^2w_{\sigma } =\frac{2}{a M_P} \left[ E_0 \left( \frac{a'}{a} u_{\sigma } - u_{\sigma }^{\prime} \right) - \sigma k B_0u_{\sigma } \right].  
\end{align}
Let $u_k(\tau) = \mathrm{Re}\lb U_k(\tau) e^{ik(\tau-\tau_m)} \rb$ and likewise for $w$. Here $\tau_m \sim 10^4/k$ is a constant matching time at which we match the high frequency solution to the numerical solution. Dropping subscripts for now, the equations of motion become 
\begin{eqnarray}
 U\prpr +2ik U\pr - \frac{a\prpr}{a}U +2 \frac{B_0^2 - E_0^2}{M_{P}^2 a^2}U = \frac{2}{a M_P} \lb E_0 W\pr + ikE_0W - \sigma k B_0 W\rb \\
    W\prpr +2ik W\pr = \frac{2}{a M_P} \lb -E_0 U\pr -ik E_0 U +E_0 \mch U - \sigma k B_0 U \rb .
\end{eqnarray}
We are searching for a high-frequency solution, so assume the following four conditions:  $ k \left|U \right| \gg  \left| U\pr \right|$, $ k \gg \mch $, $  k \left| U\pr \right| \gg \left| (a\prpr/a) U \right|$, and $k \left| U\pr \right| \gg \left|U (B_0^2 -E_0^2)/(a^2 M_{P}^2) \right| $. The first condition assumes the envelope functions $U,W$ vary appreciably only on time scales much larger than $1/k$. The second condition assumes the mode is deep within the horizon. The third condition is easily satisfied during radiation domination, during which the gauge radiation terms are most significant. Using the definition in Eq.~(\ref{eqn:RERBdefn}), the last condition can be expressed 
\beq
 \frac{a^2 }{2 \left|R_B - R_E \right| \Omega_{SMR} H_0^2} \frac{k\left| U\pr \right|}{\left| U \right|}  \gg 1 \label{fieldterm}
\eeq
which we will show shortly is valid in the regime of interest. When the above four conditions hold the equations of motion simplify to
\begin{align}
    U\pr  &= \frac{1}{a M_P} \lb  E_0 W + i \sigma  B_0 W\rb \label{UprimeEOM} \\
    W\pr &= -\frac{1}{a M_P} \lb  E_0 U  - i \sigma  B_0 U \rb .
\end{align}
Defining a new time variable $dx \equiv d\tau / a$ then yields equations of motion for simple harmonic oscillators in $x$, giving
\beq
U = U_+ e^{i \omega x} + U_-  e^{-i \omega x} \label{eqn:highfreqenv}
\eeq
where $\omega = \sqrt{(E_0^2 + B_0^2)/M_P^2}$. We can now return to the condition in Eq.~(\ref{fieldterm}) which becomes
\beq
\frac{ak}{\Omega_{SMR}^{1/2}H_0}\frac{\sqrt{2(R_B+R_E)}}{2\left| R_B-R_E \right|} \frac{|U_+ e^{i\omega x} - U_- e^{-i\omega x}|}{|U_+ e^{i\omega x} + U_- e^{-i\omega x}|} \gg 1
\eeq
which is satisfied for all relevant modes. This is easily seen by recognizing that the fraction containing $R_E,R_B$ is always larger than unity, and the fraction containing $k/H_0$ is roughly $k\tau$ during the radiation era, which is large by assumption $k\gg \mch$.

Next we detail how this analytic solution is used to compute the spectral energy densities. The high frequency solution is matched at time $\tau_m$ to the numerical solution, so it is convenient to redefine the coefficients $U_+,U_-$ so that
\beq
U = (\Upc + i \Ups) e^{-ik\tau_m - i \omega x_m} e^{i \omega x }+ (\Umc + i \Ums) e^{-ik\tau_m + i \omega x_m} e^{-i \omega x},
\eeq
where $(\Upc, \Ups, \Umc,\Ums) \in \mathbb{R}$. The above choice yields the full high frequency solution
\begin{align}
u_k(\tau) = \Upc \cos(\delta_+) - \Ups \sin(\delta_+) +\Umc \cos(\delta_-) - \Ums \sin(\delta_-)
\end{align}
where we have defined the argument $\delta_{\pm} \equiv k(\tau-\tau_m) \pm \omega (x - x_m)$,  so
\beq
\frac{d\delta_{\pm}}{d\tau} = k \pm  \frac{\omega}{a(\tau)} \equiv k \pm f (\tau).
\eeq
Note $k\gg \mch$ implies $k \gg f$. The solution for $W$ is obtained with Eq.~(\ref{UprimeEOM}) and inserting the solution for $U$, giving
\begin{align}
    w = W_{c+} \cos(\delta_+) - W_{s+} \sin(\delta_+) + W_{c-} \cos(\delta_-) -W_{s-} \sin(\delta_-)
\end{align}
with the $W_{s,c,\pm}$ coefficients
\begin{align}
W_{c+} &\equiv  \frac{\sigma B_0 \Upc - E_0 \Ups}{M_P\omega} \\
W_{s+} &\equiv  \frac{\sigma B_0 \Ups + E_0 \Upc}{M_P \omega} \\
W_{c-} &\equiv - \frac{(\sigma B_0 \Umc -  E_0 \Ums)}{M_P \omega} \\
W_{s-} &\equiv  -\frac{(\sigma B_0 \Ums + E_0 \Umc)}{M_P \omega}.
\end{align}
The four independent coefficients then are found by solving a system of four linear equations 
\begin{align}
    u_k(\tau = \tau_m ) &\equiv u_m  = \Upc + \Umc \\
u_k\pr(\tau= \tau_m) & \equiv du_m   = -(k + f_m)\Ups - (k-f_m)\Ums\\
  w_k(\tau = \tau_m ) &\equiv w_m  = \frac{(\sigma B_0 \Upc - E_0 \Ups)}{M_P \omega} - \frac{(\sigma B_0 \Umc - E_0 \Ums)}{M_P \omega} \\
w_k\pr(\tau= \tau_m) & \equiv dw_m   = -(k + f_m)\frac{(\sigma B_0 \Ups + E_0 \Upc)}{M_P \omega} + (k-f_m)\frac{(\sigma B_0 \Ums + E_0 \Umc)}{M_P \omega}
\end{align}
for $f_m \equiv f(\tau_m)$, with solutions
\begin{align}
    \Upc &= \frac{-\bar{E}_0 k \omega dw_m + \bar{E}_0^2 k(k-f_m) u_m + \sigma \bar{B}_0 \bar{E}_0 f_m du_m + (k^2 - f_m^2)\sigma \bar{B}_0(\omega w_m + \sigma \bar{B}_0 u_m)}{2k^2 \omega^2 - 2f_m^2 \bar{B}_0^2} \label{Upceqn}\\
    \Ups &= \frac{(f_m-k) \lb \bar{E}_0 k \omega w_m + \sigma \bar{B}_0 \omega dw_m + \sigma \bar{B}_0 \bar{E}_0 f_m u_m \rb - du_m \lb (k-f_m)\bar{B}_0^2 + k \bar{E}_0^2 \rb}{2k^2 \omega^2 - 2f_m^2 \bar{B}_0^2} \label{UpSeqn}\\
    \Umc &= \frac{\bar{E}_0 k \omega dw_m + \bar{E}_0^2 k (k+f_m) u_m - \sigma \bar{B}_0 \bar{E}_0 f_m du_m + (k^2 -f_m^2)\sigma \bar{B}_0 \lb \sigma \bar{B}_0 u_m - \omega w_m \rb}{2k^2 \omega^2 - 2f_m^2 \bar{B}_0^2} \label{UMceqn}\\
    \Ums &= \frac{- \lb \bar{E}_0^2 k + (k+f_m)\bar{B}_0^2 \rb du_m + (k+f_m) \lb \bar{E}_0 k \omega w_m + \sigma \bar{B}_0 \omega dw_m + \sigma \bar{B}_0 \bar{E}_0 f_m u_m \rb}{2k^2 \omega^2 - 2f_m^2 \bar{B}_0^2} , \label{Umseqn}
\end{align}
where $(\bar{E}_0,\bar{B}_0) \equiv \tfrac{1}{M_P}(E_0, B_0)$ have been defined for notational compactness. 

For the purposes of computing the spectral energy densities, we need the time average of $u\pr{}^2$ (evaluated at $\tau_0$); we have
\beq
u\pr_k = -(k+f)\lb \Upc \sin (\delta_+) + \Ups \cos(\delta_+)  \rb - (k-f) \lb \Umc \sin(\delta_-) + \Ums \cos(\delta_-) \rb.
\eeq
The trigonometric functions in $\delta_{\pm}$ can be expanded into products of trigonometric functions in $k\tau$ and $\omega x$. For all modes of interest we have $k \gg f$ and $k\gg \mch$, so  $\omega x$ and $f(\tau)$ are effectively constant over the averaging period $2\pi /k $. The time average can then be safely taken only over the functions in $k\tau$, giving
\begin{align}
\langle u\pr_k{}^2\rangle_{\tau} = \frac{1}{2} &\biggl\{ \bigl[ k^2((\Upc+\Umc)^2+(\Ups + \Ums)^2)  +2fk (\Upc^2 + \Ups^2 - \Umc^2 - \Ums^2)\nonumber \\
&+ f^2 ((\Upc-\Umc)^2+(\Ups - \Ums)^2) \bigr]  \cos^2 \lb \omega (x-x_m)\rb \nonumber \\
&+ \bigl[ k^2 ((\Upc-\Umc)^2+(\Ups - \Ums)^2) +2fk (\Upc^2 + \Ups^2 - \Umc^2 - \Ums^2)\nonumber \\
&+ f^2 ((\Upc+\Umc)^2+(\Ups + \Ums)^2)) \bigr]  \sin^2 \lb \omega (x-x_m)\rb \nonumber \\
&+  4 (k^2 - f^2)(\Upc \Ums - \Ups \Umc)\sin \lb \omega (x-x_m)\rb \cos \lb \omega (x-x_m)\rb
\biggr\} \label{eqn:uprimeavg}
\end{align}
and an identical expression for $\langle w\pr_k{}^2 \rangle$ with the replacement $\left\{\Upc, \Ups, \Umc,\Ums \right\} \rightarrow \left\{W_{c+},W_{s+},W_{c-},W_{s-} \right \}$. Eq.~(\ref{eqn:uprimeavg}) and the $w$ analogue are what we use in practice to compute the spectral energy densities $\Omega_{GW}$ and $\Omega_{\delta A}$ in Eqs.~(\ref{eqn:OmegaGW}) and (\ref{eqn:OmegadeltaA}) respectively. The trigonometric functions in $\omega x$ that appear in Eq.~(\ref{eqn:uprimeavg}) are the cause of the oscillations described in Sec.~\ref{subsec:oscillations}, and for a mode that enters in the radiation dominated era, $x-x_m$ has a $\ln(k)$ contribution. These provide the justification for the approximate fitting formula in Eq.~(\ref{eqn:oscfitformula}), although it is clear that the frequency dependence of Eq.~(\ref{eqn:uprimeavg}) is complicated.

\section{Modified Superhorizon GW Evolution}
\label{AppendixSuperhorizon}

In this appendix we solve the equations of motion (\ref{eqn:eomu}-\ref{eqn:eomw}) in the low-frequency limit to examine the effect of the U(1) fields on superhorizon GW evolution. In the $k=0$ limit the equations of motion become
\begin{eqnarray}
u \prpr  - \frac{a\prpr}{a}u +2 \frac{B_0^2 - E_0^2}{M_{pl}^2 a^2} u \label{ulowkeom}  = \frac{2E_0}{a M_{P}} \lb  w\pr   \rb\\
w \prpr = \frac{2E_0}{a M_{P}}\lb - u\pr + \mch u \label{wlowkeom}  \rb.
\end{eqnarray}
Dividing both equations by $u_{k,\mathrm{prim}}$ recasts the equations of motion in term of the transfer functions $\mct_u,\mct_w$. We again focus on the radiation era during which the background U(1) terms present here are most relevant, so $a(\tau) = a_{RH}(\tau/\tau_{RH}) = \Omega_R^{1/2} H_0 \tau$ (see Section \ref{subsec:init}). Then the equations of motion can be compactly written,
\begin{eqnarray}
\mct_u \prpr  +2 \frac{\bar{B}_0^2 - \bar{E}_0^2}{\Omega_R H_0^2 \tau^{2}} \mct_u   = \frac{2\bar{E}_0}{\Omega_R^{1/2} H_0\tau} \lb  \mct_w\pr   \rb\\
\mct_w \prpr = -\frac{2\bar{E}_0}{\Omega_R^{1/2} H_0} \frac{d}{d\tau}\lb \mct_u \tau^{-1} \rb,
\end{eqnarray}
which are easily decoupled and solved by integrating the second equation of motion and inserting it into the first, keeping in mind the initial conditions in Eq.~(\ref{eqn:TuTwICs}). This gives the $\mct_u$ equation of motion 
\beq
\tau_{RH}^2\mct_u \prpr + 2\frac{(\bar{B}_0^2 + \bar{E}_0^2)}{ \Omega_R H_0^2 (\tau/\tau_{RH})^{2}} \mct_u   - \frac{4 \bar{E}_0^2}{\Omega_R H_0^2 (\tau/\tau_{RH})}  =0.
\eeq
Dropping the negligible $\mathcal{O}(R_X^2)$ difference between $(E_0^2,B_0^2)/\Omega_{R}$ and $(E_0^2,B_0^2)/\Omega_{SMR}$, this yields simple solutions,
\begin{align}
\mct_u &= \mathcal{B}_R \lp \frac{\tau}{\tau_{RH}} \rp^{n_{-}} + \mathcal{A}_R \lp \frac{\tau}{\tau_{RH}} \rp^{n_{+}} + \frac{2R_E}{R_B+R_E}\frac{\tau}{\tau_{RH}}\\
\mct_w &= -2\sqrt{2R_E}\left[  \frac{R_B-R_E}{R_B+R_E}\lp \frac{\tau}{\tau_{RH}}-1\rp + \frac{\mathcal{B}_R}{n_-} \left( \frac{\tau}{\tau_{RH}} \right)^{n_-} + \frac{\mathcal{A}_R}{n_+} \left( \frac{\tau}{\tau_{RH}} \right)^{n_+} \right]
\end{align}
for $n_{\pm} = \frac{1}{2} \Bigl(1 \pm  \sqrt{1- 16(R_B+R_E)} \Bigr)$ and
\begin{align}
\mathcal{B}_R =\frac{R_E - R_B}{R_B + R_E}  \frac{1-\sqrt{1-16(R_B+R_E)}}{2\sqrt{1-16(R_B+R_E)}} &=\frac{R_E - R_B}{R_B + R_E}  \frac{n_-}{\sqrt{1-16(R_B+R_E)}} \\
\mathcal{A}_R = \frac{R_B - R_E}{R_B + R_E}  \frac{1+\sqrt{1-16(R_B+R_E)}}{2\sqrt{1-16(R_B+R_E)}} &= \frac{R_B - R_E}{R_B + R_E}  \frac{n_+}{\sqrt{1-16(R_B+R_E)}},
\end{align}
where $|\mathcal{A}_R|>|\mathcal{B}_R|$. In the standard case without the gauge fields, the GW transfer function equation of motion (also for $k\rightarrow 0$ and in the radiation era) is $\mct_{u,\rm{std}} \prpr =0$ with solution $\mct_{u,\rm{std}} =\tau/\tau_{RH}$. Therefore in the presence of the gauge fields the GW amplitude is altered while outside the horizon by a factor 
\beq
\frac{\mct_u(\tau)}{\mct_{u,\rm{std}}(\tau)} = \mathcal{B}_R \lp \frac{\tau}{\tau_{RH}} \rp ^{\frac{1}{2}\lp-1  - \sqrt{1-16(R_B+R_E)} \rp} + \mathcal{A}_R \lp \frac{\tau}{\tau_{RH}} \rp ^{\frac{1}{2}\lp -1+ \sqrt{1-16(R_B+R_E)} \rp} + \frac{2R_E}{R_B+R_E}
\eeq
before the mode enters the horizon at $\tau_{entry} = 1/k$. Lower frequency modes enter later and accumulate a larger modification, so this effect is frequency dependent. Taking a piecewise approximation in which this $k\rightarrow 0$ description is valid until $\tau=\tau_{entry}$, after which point the amplitude modulation shuts off and subhorizon oscillation begins, this modulated amplitude means that $\Omega_{GW}$ is different by a factor
\begin{align}
\left| \mct_{u_{k}}(\tau_{entry}) / \mct_{u_{k},\rm{std}}(\tau_{entry}) \right|^2 = \biggl[ & \mathcal{B}_R \lp \frac{k}{k_{RH}}\rp^{\frac{1}{2}\lp 1 + \sqrt{1-16(R_B+R_E)} \rp} + \mathcal{A}_R \lp \frac{k}{k_{RH}}\rp^{\frac{1}{2}\lp 1- \sqrt{1-16(R_B+R_E)} \rp}  + \frac{2R_E}{R_B+R_E} \biggr]^2
\end{align}
where we have restored the $k$-subscript and used $\tau_{RH} = 1/k_{RH}$. The power laws in $k$ give the GW spectrum the new tilt seen in this model. When the background fields are absent, the standard case is recovered, and there is no additional tilt. The tilt also disappears for $R_E\approx R_B$ so it is useful to consider the limiting cases when one background field is zero,
\begin{align}
P_B &= \left| \mct_{u_{k}}(\tau_{entry}) / \mct_{u_{k},\rm{std}}(\tau_{entry}) \right|^2_{R_E=0} = \biggl[  \mathcal{B}_R \lp \frac{k}{k_{RH}}\rp^{\frac{1}{2}\lp 1 + \sqrt{1-16 R_B }\rp} + \mathcal{A}_R \lp \frac{k}{k_{RH}}\rp^{\frac{1}{2}\lp 1- \sqrt{1-16 R_B} \rp}\biggr]^2 \\
P_E &= \left| \mct_{u_{k}}(\tau_{entry}) / \mct_{u_{k},\rm{std}}(\tau_{entry}) \right|^2_{R_B=0} = \biggl[ 2+ \mathcal{B}_R \lp \frac{k}{k_{RH}}\rp^{\frac{1}{2}\lp 1 + \sqrt{1-16 R_E} \rp} + \mathcal{A}_R \lp \frac{k}{k_{RH}}\rp^{\frac{1}{2}\lp 1- \sqrt{1-16 R_E} \rp}\biggr]^2.
\end{align}
We always have $k<k_{RH}$, so the $\mathcal{A}_R$ term is more significant; dropping the subdominant $\mathcal{B}_R$ term then gives  
\begin{align}
P_B &\simeq \biggl[ \mathcal{A}_R \lp \frac{k}{k_{RH}}\rp^{\frac{1}{2}\lp 1- \sqrt{1-16R_B} \rp} \biggr]^2 \label{eqn:PBeqn} \\
P_E &\simeq  \biggl[ 2 + \mathcal{A}_R \lp \frac{k}{k_{RH}}\rp^{\frac{1}{2}\lp 1- \sqrt{1-16 R_E} \rp} \biggr]^2 \label{eqn:PEeqn}.
\end{align}
Hence at low frequencies the E-enhancement is limited to roughly a factor of 4 while the B-suppression can be far more significant. Once radiation domination gives way to matter domination at $\tau_{eq}$, the $E_0,B_0$ background rapidly dilutes away and the modification to the superhorizon evolution ends. Thus all modes $k\lesssim k_{eq}$ accumulate the same maximum modification. This is why the tilt for $k\lesssim k_{eq}$ becomes a constant offset. Hence we replace $k\rightarrow \mathrm{min}(k,k_{eq})$ in Eqs.~(\ref{eqn:PBeqn}-\ref{eqn:PEeqn}).

\section{Sextet Model}
\label{AppendixSixU1}
In this section we extend the model to incorporate six U(1) fields, in which each field obtains only a single electric or magnetic background. The corresponding ansatz is
%
\begin{equation}
 F^{(i)}_{j0} = E_0 \delta^i_j, \qquad F^{(i')}_{k j} = B_0 \epsilon^{(i'-3)jk}
\end{equation}
where charge flavor $i=1,2,3$ is mapped to Cartesian directions $x,y,z$, whereas charge flavor $i'=4,5,6$ is separately mapped to $x,y,z$. That is, the electric- and magnetic-like fields are associated with different triplets of U(1) fields. This field configuration gives identical background behavior to the triplet case, but the linear wavelike perturbations 
\begin{align}
     \delta A_{\mu}^{(1)} &= \frac{M_P}{2}\left(0,w_+(\tau,z),w_{\times}(\tau,z),0 \right)\\
     \delta A_{\mu}^{(2)} &= \frac{M_P}{2}\left(0,w_{\times}(\tau,z),-w_{+}(\tau,z),0 \right)\\
     \delta A_{\mu}^{(4)} &= \frac{M_P}{2}\left(0,y_+(\tau,z),y_{\times}(\tau,z),0 \right)\\
     \delta A_{\mu}^{(5)} &= \frac{M_P}{2}\left(0,y_{\times}(\tau,z),-y_{+}(\tau,z),0 \right)
\end{align}
change the gravitational wave dynamics. The equations of motion
\begin{eqnarray}
u_{\sigma}^{\prime \prime}  +  \left( k^2 - \frac{a^{\prime \prime}}{a} +2 \frac{(B_0^2-E_0^2)}{a^2 M_P^2 }\right) u_{\sigma } = \frac{2}{a M_P} \left[ E_0 w_{\sigma }^{\prime} - \sigma k B_0  y_{\sigma } \right] \\
w_{\sigma }^{\prime \prime} + k^2w_{\sigma } =\frac{2}{a M_P} \left[ E_0 \left( \frac{a'}{a} u_{\sigma } - u_{\sigma }^{\prime} \right) \right]\\
y_{\sigma }^{\prime \prime} + k^2y_{\sigma } =\frac{2}{a M_P} \left[  - \sigma k B_0u_{\sigma } \right]
\end{eqnarray}
demonstrate different features than the triplet case. The equations of motion are invariant under $(\sigma,y)\rightarrow -(\sigma,y)$ or $(\sigma,u,w)\rightarrow -(\sigma,u,w)$, neither of which modify the spectral energy densities. Hence the spectral energy densities do not exhibit an excess handedness. Furthermore, the sextet and triplet models have the same equations of motion the limit that either $E_0$ or $B_0$ vanish, and in the superhorizon $k\rightarrow 0$ limit.

An analogous high frequency solution exists for this system, giving the following system, 
\begin{align}
    \frac{dU}{dx} &= \bar{E}_0 W + i \sigma \bar{B}_0 Y \\
    \frac{dW}{dx} & = -\bar{E}_0 U \\
    \frac{dY}{dx} & = i \sigma \bar{B}_0 U.
\end{align}
The GW equation of motion can again be cast in terms of a harmonic oscillator in $x$ with frequency $\omega$, but the gauge field system is changed,
\begin{align}
    U &= U_+ e^{i\omega x} + U_- e^{-i\omega x} \\
    W &= W_0 e^{-ik\tau_m} + i \frac{\bar{E}_0}{\omega} U_+ e^{i\omega x} - i \frac{\bar{E}_0}{\omega} U_- e^{-i\omega x} \\ 
    Y &=i \frac{\bar{E}_0}{\sigma \bar{B}_0} W_0 e^{-ik\tau_m} + \sigma \frac{\bar{B}_0}{\omega} U_+ e^{i\omega x} -  \sigma \frac{\bar{B}_0}{\omega} U_- e^{-i\omega x}
\end{align}
where we have three complex coefficients for our six initial conditions. The corresponding solution to the coupled system is
\begin{align}
    u &= \Upc \cos(\delta_+) - \Ups \sin(\delta_+) + \Umc \cos(\delta_-) - \Ums \sin(\delta_-) \\
    w &= W_1 \cos(k(\tau-\tau_m)) - W_2 \sin(k(\tau-\tau_m)) + \frac{\bar{E}_0}{\omega} \left[ -\Upc \sin(\delta_+) - \Ups \cos(\delta_+) + \Umc \sin(\delta_-) +\Ums \cos(\delta_-) \right] \\ 
    y &= \frac{\bar{E}_0 }{\sigma \bar{B}_0} \left[ -W_2 \cos(k(\tau-\tau_m)) - W_1 \sin(k(\tau-\tau_m)) \right] + \frac{\sigma \bar{B}_0 }{\omega} \left[ \Upc \cos(\delta_+) - \Ups \sin(\delta_+) - \Umc \cos(\delta_-) +\Ums \sin(\delta_-) \right]
\end{align}
which yields the following four matching conditions
\begin{align}
    u_m &= \Upc + \Umc \\
    du_m &= -\Ups(k+f_m) - \Ums (k-f_m)\\
    w_m &= W_1 + \frac{\bar{E}_0}{\omega} \left[\Ums - \Ups \right] \\
    dw_m &= -kW_2 + \frac{\bar{E}_0}{\omega} \left[ -\Upc (k+f_m) + \Umc (k-f_m) \right] \\ 
    y_m &= -\frac{\bar{E}_0}{\sigma \bar{B}_0 } W_2 + \frac{\sigma \bar{B}_0}{\omega} \left[ \Upc - \Umc \right] \\
    dy_m &= -\frac{\bar{E}_0}{\sigma \bar{B}_0 } k W_1 + \frac{\sigma \bar{B}_0}{\omega} \left[ -\Ups(k+f_m) + \Ums(k-f_m) \right] 
\end{align}
whose solutions give the coefficients 
\begin{align}
    \Upc &= \frac{\bar{E}_0^2 (k-f_m)u_m - \omega\bar{E}_0 dw_m + k \bar{B}_0 \left( \bar{B}_0 u_m + \sigma \omega y_m \right)}{2k \omega^2} \\
    \Umc &= \frac{\bar{E}_0^2 (k+f_m)u_m + \omega\bar{E_0} dw_m + k \bar{B}_0 \left( \bar{B}_0 u_m - \sigma \omega y_m \right)}{2k \omega^2} \\ 
    \Ups &= \frac{-\bar{B}_0^2(k-f_m)du_m - k\bar{E}_0^2 du_m - \omega(k-f_m) (k \bar{E}_0 w_m + \sigma \bar{B}_0 dy_m) }{2k^2\bar{E}_0^2 +2\bar{B}_0^2(k^2-f_m^2) }\\ 
    \Ums &= \frac{-\bar{B}_0^2(k+f_m)du_m - k\bar{E}_0^2 du_m + \omega (k+f_m)(k \bar{E}_0 w_m + \sigma \bar{B}_0 dy_m) }{2k^2\bar{E}_0^2 +2\bar{B}_0^2(k^2-f_m^2) } \\
    W_1 &= \frac{\bar{B}_0 \left[-\sigma k \bar{E}_0\omega dy_m + f_m\bar{B}_0\bar{E}_0 du_m + (k^2 - f_m^2)\omega \bar{B}_0 w_m\right] }{ \omega \left( k^2 \bar{E}_0^2 + (k^2 - f_m^2)\bar{B}_0^2 \right)} \\
    W_2 &= \frac{-\bar{B}_0 \left[ \sigma k \bar{E}_0 \omega y_m + f_m\bar{B}_0\bar{E}_0 u_m + \omega \bar{B}_0 dw_m \right] }{ k\omega^3}.
\end{align}
For computing the spectral energy densities, we still have Eq. (\ref{eqn:uprimeavg}) for the GW, although the coefficients $\Upc$, etc. have changed. The spectral energy densities for the gauge field system, on the other hand, are slightly modified. It is convenient to define a new set of coefficients,
\begin{align}
    W_{+c} &= -\frac{\bar{E}_0}{\omega}U_{+s} \qquad W_{+s} = \frac{\bar{E}_0}{\omega}U_{+c} \\
    W_{-c} &= \frac{\bar{E}_0}{\omega}U_{-s}  \qquad W_{-s} = - \frac{\bar{E}_0}{\omega}U_{-c} \nonumber \\
    Y_{+c} &= \frac{\sigma \bar{B}_0}{\omega}U_{+c} \qquad Y_{+s} = \frac{\sigma \bar{B}_0}{\omega}U_{+s} \nonumber \\
    Y_{-c} &= -\frac{\sigma \bar{B}_0}{\omega}U_{-c}  \qquad Y_{-s} = - \frac{\sigma \bar{B}_0}{\omega}U_{-s} \nonumber \\
    Y_1 &= -\frac{\bar{E}_0}{\sigma \bar{B}_0}W_2 \qquad  Y_2 = \frac{\bar{E}_0}{\sigma \bar{B}_0}W_1 \nonumber
\end{align}
which casts the solutions into similar forms,
\begin{align}
    u &= \Upc \cos(\delta_+) - \Ups \sin(\delta_+) + \Umc \cos(\delta_-) - \Ums \sin(\delta_-) \\
    w &= W_1 \cos(k(\tau-\tau_m)) - W_2 \sin(k(\tau-\tau_m)) +  W_{+c} \cos(\delta_+) - W_{+s} \sin(\delta_+) +  W_{-c} \cos(\delta_-) - W_{-s} \sin(\delta_-) \\ 
    y &=  Y_1 \cos(k(\tau-\tau_m)) - Y_2 \sin(k(\tau-\tau_m)) +   Y_{+c} \cos(\delta_+) - Y_{+s} \sin(\delta_+) +  Y_{-c} \cos(\delta_-) - Y_{-s} \sin(\delta_-).
\end{align}
These resemble the solutions from the triplet scenario, except the gauge field excitations have acquired homogeneous solutions. These homogeneous solutions contribute additional terms to the time average of the squared  gauge field mode functions, 
\begin{align}
    \langle {w'}_k^2\rangle =& \langle {w'}_u^2 \rangle + \frac{k^2(W_1^2 + W_2^2)}{2}  +2\frac{kW_1}{2} \bigl[(k+f)W_{+c} \cos(\omega(x-x_m)) -(k+f)W_{+s} \sin(\omega(x-x_m)) \nonumber \\ 
    &+ (k-f)W_{-c}\cos(\omega(x-x_m)) 
    +(k-f) W_{-s}\sin(\omega(x-x_m)) \bigr]
   +2\frac{kW_2}{2}\bigl[(k+f)W_{+c} \sin(\omega(x-x_m)) \nonumber \\
   &+ (k+f)W_{+s} \cos(\omega(x-x_m)) -(k-f)W_{-c}\sin(\omega(x-x_m)) +(k-f) W_{-s}\cos(\omega(x-x_m)) \bigr]
\end{align}
where $\langle w_u'^2 \rangle$ schematically indicates usage of the expression for $u$, Eqn.~(\ref{eqn:uprimeavg}), with the coefficient replacement $\left\{\Upc, \Ups, \Umc,\Ums \right\} \rightarrow \left\{W_{c+},W_{s+},W_{c-},W_{s-} \right \}$.

\end{widetext}
\vfill
\eject

\bibliography{main}

\begin{thebibliography}{94}
\expandafter\ifx\csname natexlab\endcsname\relax\def\natexlab#1{#1}\fi
\expandafter\ifx\csname bibnamefont\endcsname\relax
  \def\bibnamefont#1{#1}\fi
\expandafter\ifx\csname bibfnamefont\endcsname\relax
  \def\bibfnamefont#1{#1}\fi
\expandafter\ifx\csname citenamefont\endcsname\relax
  \def\citenamefont#1{#1}\fi
\expandafter\ifx\csname url\endcsname\relax
  \def\url#1{\texttt{#1}}\fi
\expandafter\ifx\csname urlprefix\endcsname\relax\def\urlprefix{URL }\fi
\providecommand{\bibinfo}[2]{#2}
\providecommand{\eprint}[2][]{\url{#2}}

\bibitem[{\citenamefont{Guth}(1981)}]{Guth:1980zm}
\bibinfo{author}{\bibfnamefont{A.~H.} \bibnamefont{Guth}},
  \bibinfo{journal}{Phys. Rev. D} \textbf{\bibinfo{volume}{23}},
  \bibinfo{pages}{347} (\bibinfo{year}{1981}).

\bibitem[{\citenamefont{Linde}(1982)}]{Linde:1981mu}
\bibinfo{author}{\bibfnamefont{A.~D.} \bibnamefont{Linde}},
  \bibinfo{journal}{Phys. Lett. B} \textbf{\bibinfo{volume}{108}},
  \bibinfo{pages}{389} (\bibinfo{year}{1982}).

\bibitem[{\citenamefont{Albrecht and Steinhardt}(1982)}]{Albrecht:1982wi}
\bibinfo{author}{\bibfnamefont{A.}~\bibnamefont{Albrecht}} \bibnamefont{and}
  \bibinfo{author}{\bibfnamefont{P.~J.} \bibnamefont{Steinhardt}},
  \bibinfo{journal}{Phys. Rev. Lett.} \textbf{\bibinfo{volume}{48}},
  \bibinfo{pages}{1220} (\bibinfo{year}{1982}).

\bibitem[{\citenamefont{Mukhanov and Chibisov}(1981)}]{Mukhanov:1981xt}
\bibinfo{author}{\bibfnamefont{V.~F.} \bibnamefont{Mukhanov}} \bibnamefont{and}
  \bibinfo{author}{\bibfnamefont{G.~V.} \bibnamefont{Chibisov}},
  \bibinfo{journal}{JETP Lett.} \textbf{\bibinfo{volume}{33}},
  \bibinfo{pages}{532} (\bibinfo{year}{1981}).

\bibitem[{\citenamefont{Hawking}(1982)}]{Hawking:1982cz}
\bibinfo{author}{\bibfnamefont{S.~W.} \bibnamefont{Hawking}},
  \bibinfo{journal}{Phys. Lett. B} \textbf{\bibinfo{volume}{115}},
  \bibinfo{pages}{295} (\bibinfo{year}{1982}).

\bibitem[{\citenamefont{Guth and Pi}(1982)}]{Guth:1982ec}
\bibinfo{author}{\bibfnamefont{A.~H.} \bibnamefont{Guth}} \bibnamefont{and}
  \bibinfo{author}{\bibfnamefont{S.~Y.} \bibnamefont{Pi}},
  \bibinfo{journal}{Phys. Rev. Lett.} \textbf{\bibinfo{volume}{49}},
  \bibinfo{pages}{1110} (\bibinfo{year}{1982}).

\bibitem[{\citenamefont{Starobinsky}(1982)}]{Starobinsky:1982ee}
\bibinfo{author}{\bibfnamefont{A.~A.} \bibnamefont{Starobinsky}},
  \bibinfo{journal}{Phys. Lett. B} \textbf{\bibinfo{volume}{117}},
  \bibinfo{pages}{175} (\bibinfo{year}{1982}).

\bibitem[{\citenamefont{Bardeen et~al.}(1983)\citenamefont{Bardeen, Steinhardt,
  and Turner}}]{Bardeen:1983qw}
\bibinfo{author}{\bibfnamefont{J.~M.} \bibnamefont{Bardeen}},
  \bibinfo{author}{\bibfnamefont{P.~J.} \bibnamefont{Steinhardt}},
  \bibnamefont{and} \bibinfo{author}{\bibfnamefont{M.~S.}
  \bibnamefont{Turner}}, \bibinfo{journal}{Phys. Rev. D}
  \textbf{\bibinfo{volume}{28}}, \bibinfo{pages}{679} (\bibinfo{year}{1983}).

\bibitem[{\citenamefont{Grishchuk}(1975)}]{Grishchuk:1974ny}
\bibinfo{author}{\bibfnamefont{L.~P.} \bibnamefont{Grishchuk}},
  \bibinfo{journal}{Sov. Phys. JETP} \textbf{\bibinfo{volume}{40}},
  \bibinfo{pages}{409} (\bibinfo{year}{1975}).

\bibitem[{\citenamefont{Starobinsky}(1979)}]{Starobinsky:1979ty}
\bibinfo{author}{\bibfnamefont{A.~A.} \bibnamefont{Starobinsky}},
  \bibinfo{journal}{JETP Lett.} \textbf{\bibinfo{volume}{30}},
  \bibinfo{pages}{682} (\bibinfo{year}{1979}).

\bibitem[{\citenamefont{Rubakov et~al.}(1982)\citenamefont{Rubakov, Sazhin, and
  Veryaskin}}]{Rubakov:1982df}
\bibinfo{author}{\bibfnamefont{V.~A.} \bibnamefont{Rubakov}},
  \bibinfo{author}{\bibfnamefont{M.~V.} \bibnamefont{Sazhin}},
  \bibnamefont{and} \bibinfo{author}{\bibfnamefont{A.~V.}
  \bibnamefont{Veryaskin}}, \bibinfo{journal}{Phys. Lett. B}
  \textbf{\bibinfo{volume}{115}}, \bibinfo{pages}{189} (\bibinfo{year}{1982}).

\bibitem[{\citenamefont{Fabbri and Pollock}(1983)}]{Fabbri:1983us}
\bibinfo{author}{\bibfnamefont{R.}~\bibnamefont{Fabbri}} \bibnamefont{and}
  \bibinfo{author}{\bibfnamefont{M.~d.} \bibnamefont{Pollock}},
  \bibinfo{journal}{Phys. Lett. B} \textbf{\bibinfo{volume}{125}},
  \bibinfo{pages}{445} (\bibinfo{year}{1983}).

\bibitem[{\citenamefont{Abbott and Wise}(1984)}]{Abbott:1984fp}
\bibinfo{author}{\bibfnamefont{L.~F.} \bibnamefont{Abbott}} \bibnamefont{and}
  \bibinfo{author}{\bibfnamefont{M.~B.} \bibnamefont{Wise}},
  \bibinfo{journal}{Nucl. Phys. B} \textbf{\bibinfo{volume}{244}},
  \bibinfo{pages}{541} (\bibinfo{year}{1984}).

\bibitem[{\citenamefont{Kamionkowski et~al.}(1997)\citenamefont{Kamionkowski,
  Kosowsky, and Stebbins}}]{Kamionkowski:1996zd}
\bibinfo{author}{\bibfnamefont{M.}~\bibnamefont{Kamionkowski}},
  \bibinfo{author}{\bibfnamefont{A.}~\bibnamefont{Kosowsky}}, \bibnamefont{and}
  \bibinfo{author}{\bibfnamefont{A.}~\bibnamefont{Stebbins}},
  \bibinfo{journal}{Phys. Rev. Lett.} \textbf{\bibinfo{volume}{78}},
  \bibinfo{pages}{2058} (\bibinfo{year}{1997}), \eprint{astro-ph/9609132}.

\bibitem[{\citenamefont{Seljak and Zaldarriaga}(1997)}]{Seljak:1996gy}
\bibinfo{author}{\bibfnamefont{U.}~\bibnamefont{Seljak}} \bibnamefont{and}
  \bibinfo{author}{\bibfnamefont{M.}~\bibnamefont{Zaldarriaga}},
  \bibinfo{journal}{Phys. Rev. Lett.} \textbf{\bibinfo{volume}{78}},
  \bibinfo{pages}{2054} (\bibinfo{year}{1997}), \eprint{astro-ph/9609169}.

\bibitem[{\citenamefont{Ade et~al.}(2017)}]{Ade:2017uvt}
\bibinfo{author}{\bibfnamefont{P.~A.~R.} \bibnamefont{Ade}}
  \bibnamefont{et~al.} (\bibinfo{collaboration}{POLARBEAR}),
  \bibinfo{journal}{Astrophys. J.} \textbf{\bibinfo{volume}{848}},
  \bibinfo{pages}{121} (\bibinfo{year}{2017}), \eprint{1705.02907}.

\bibitem[{\citenamefont{Ade et~al.}(2018)}]{Ade:2018gkx}
\bibinfo{author}{\bibfnamefont{P.~A.~R.} \bibnamefont{Ade}}
  \bibnamefont{et~al.} (\bibinfo{collaboration}{BICEP2, Keck Array}),
  \bibinfo{journal}{Phys. Rev. Lett.} \textbf{\bibinfo{volume}{121}},
  \bibinfo{pages}{221301} (\bibinfo{year}{2018}), \eprint{1810.05216}.

\bibitem[{\citenamefont{Ade et~al.}(2019)}]{Ade:2018sbj}
\bibinfo{author}{\bibfnamefont{P.}~\bibnamefont{Ade}} \bibnamefont{et~al.}
  (\bibinfo{collaboration}{Simons Observatory}), \bibinfo{journal}{JCAP}
  \textbf{\bibinfo{volume}{02}}, \bibinfo{pages}{056} (\bibinfo{year}{2019}),
  \eprint{1808.07445}.

\bibitem[{\citenamefont{Akrami et~al.}(2020)}]{Akrami:2018odb}
\bibinfo{author}{\bibfnamefont{Y.}~\bibnamefont{Akrami}} \bibnamefont{et~al.}
  (\bibinfo{collaboration}{Planck}), \bibinfo{journal}{Astron. Astrophys.}
  \textbf{\bibinfo{volume}{641}}, \bibinfo{pages}{A10} (\bibinfo{year}{2020}),
  \eprint{1807.06211}.

\bibitem[{\citenamefont{Tristram et~al.}(2020)}]{Tristram:2020wbi}
\bibinfo{author}{\bibfnamefont{M.}~\bibnamefont{Tristram}} \bibnamefont{et~al.}
  (\bibinfo{year}{2020}), \eprint{2010.01139}.

\bibitem[{\citenamefont{Abazajian et~al.}(2016)}]{Abazajian:2016yjj}
\bibinfo{author}{\bibfnamefont{K.~N.} \bibnamefont{Abazajian}}
  \bibnamefont{et~al.} (\bibinfo{collaboration}{CMB-S4})
  (\bibinfo{year}{2016}), \eprint{1610.02743}.

\bibitem[{\citenamefont{Hazumi et~al.}(2019)}]{Hazumi:2019lys}
\bibinfo{author}{\bibfnamefont{M.}~\bibnamefont{Hazumi}} \bibnamefont{et~al.},
  \bibinfo{journal}{J. Low Temp. Phys.} \textbf{\bibinfo{volume}{194}},
  \bibinfo{pages}{443} (\bibinfo{year}{2019}).

\bibitem[{\citenamefont{Spergel and Zaldarriaga}(1997)}]{Spergel:1997vq}
\bibinfo{author}{\bibfnamefont{D.~N.} \bibnamefont{Spergel}} \bibnamefont{and}
  \bibinfo{author}{\bibfnamefont{M.}~\bibnamefont{Zaldarriaga}},
  \bibinfo{journal}{Phys. Rev. Lett.} \textbf{\bibinfo{volume}{79}},
  \bibinfo{pages}{2180} (\bibinfo{year}{1997}), \eprint{astro-ph/9705182}.

\bibitem[{\citenamefont{Kamionkowski and Kosowsky}(1998)}]{Kamionkowski:1997av}
\bibinfo{author}{\bibfnamefont{M.}~\bibnamefont{Kamionkowski}}
  \bibnamefont{and} \bibinfo{author}{\bibfnamefont{A.}~\bibnamefont{Kosowsky}},
  \bibinfo{journal}{Phys. Rev. D} \textbf{\bibinfo{volume}{57}},
  \bibinfo{pages}{685} (\bibinfo{year}{1998}), \eprint{astro-ph/9705219}.

\bibitem[{\citenamefont{Knox and Song}(2002)}]{Knox:2002pe}
\bibinfo{author}{\bibfnamefont{L.}~\bibnamefont{Knox}} \bibnamefont{and}
  \bibinfo{author}{\bibfnamefont{Y.-S.} \bibnamefont{Song}},
  \bibinfo{journal}{Phys. Rev. Lett.} \textbf{\bibinfo{volume}{89}},
  \bibinfo{pages}{011303} (\bibinfo{year}{2002}), \eprint{astro-ph/0202286}.

\bibitem[{\citenamefont{Baumann et~al.}(2015)\citenamefont{Baumann, Green, and
  Porto}}]{Baumann:2014cja}
\bibinfo{author}{\bibfnamefont{D.}~\bibnamefont{Baumann}},
  \bibinfo{author}{\bibfnamefont{D.}~\bibnamefont{Green}}, \bibnamefont{and}
  \bibinfo{author}{\bibfnamefont{R.~A.} \bibnamefont{Porto}},
  \bibinfo{journal}{JCAP} \textbf{\bibinfo{volume}{01}}, \bibinfo{pages}{016}
  (\bibinfo{year}{2015}), \eprint{1407.2621}.

\bibitem[{\citenamefont{Kamionkowski and Kovetz}(2016)}]{Kamionkowski:2015yta}
\bibinfo{author}{\bibfnamefont{M.}~\bibnamefont{Kamionkowski}}
  \bibnamefont{and} \bibinfo{author}{\bibfnamefont{E.~D.}
  \bibnamefont{Kovetz}}, \bibinfo{journal}{Ann. Rev. Astron. Astrophys.}
  \textbf{\bibinfo{volume}{54}}, \bibinfo{pages}{227} (\bibinfo{year}{2016}),
  \eprint{1510.06042}.

\bibitem[{\citenamefont{Lyth}(1997)}]{Lyth:1996im}
\bibinfo{author}{\bibfnamefont{D.~H.} \bibnamefont{Lyth}},
  \bibinfo{journal}{Phys. Rev. Lett.} \textbf{\bibinfo{volume}{78}},
  \bibinfo{pages}{1861} (\bibinfo{year}{1997}), \eprint{hep-ph/9606387}.

\bibitem[{\citenamefont{Anber and Sorbo}(2010)}]{Anber:2009ua}
\bibinfo{author}{\bibfnamefont{M.~M.} \bibnamefont{Anber}} \bibnamefont{and}
  \bibinfo{author}{\bibfnamefont{L.}~\bibnamefont{Sorbo}},
  \bibinfo{journal}{Phys. Rev. D} \textbf{\bibinfo{volume}{81}},
  \bibinfo{pages}{043534} (\bibinfo{year}{2010}), \eprint{0908.4089}.

\bibitem[{\citenamefont{Adshead and Wyman}(2012)}]{Adshead:2012kp}
\bibinfo{author}{\bibfnamefont{P.}~\bibnamefont{Adshead}} \bibnamefont{and}
  \bibinfo{author}{\bibfnamefont{M.}~\bibnamefont{Wyman}},
  \bibinfo{journal}{Phys. Rev. Lett.} \textbf{\bibinfo{volume}{108}},
  \bibinfo{pages}{261302} (\bibinfo{year}{2012}), \eprint{1202.2366}.

\bibitem[{\citenamefont{Maleknejad and
  Sheikh-Jabbari}(2013)}]{Maleknejad:2011jw}
\bibinfo{author}{\bibfnamefont{A.}~\bibnamefont{Maleknejad}} \bibnamefont{and}
  \bibinfo{author}{\bibfnamefont{M.~M.} \bibnamefont{Sheikh-Jabbari}},
  \bibinfo{journal}{Phys. Lett. B} \textbf{\bibinfo{volume}{723}},
  \bibinfo{pages}{224} (\bibinfo{year}{2013}), \eprint{1102.1513}.

\bibitem[{\citenamefont{Maleknejad et~al.}(2013)\citenamefont{Maleknejad,
  Sheikh-Jabbari, and Soda}}]{Maleknejad:2012fw}
\bibinfo{author}{\bibfnamefont{A.}~\bibnamefont{Maleknejad}},
  \bibinfo{author}{\bibfnamefont{M.~M.} \bibnamefont{Sheikh-Jabbari}},
  \bibnamefont{and} \bibinfo{author}{\bibfnamefont{J.}~\bibnamefont{Soda}},
  \bibinfo{journal}{Phys. Rept.} \textbf{\bibinfo{volume}{528}},
  \bibinfo{pages}{161} (\bibinfo{year}{2013}), \eprint{1212.2921}.

\bibitem[{\citenamefont{Maleknejad et~al.}(2018)\citenamefont{Maleknejad,
  Noorbala, and Sheikh-Jabbari}}]{Noorbala:2012fh}
\bibinfo{author}{\bibfnamefont{A.}~\bibnamefont{Maleknejad}},
  \bibinfo{author}{\bibfnamefont{M.}~\bibnamefont{Noorbala}}, \bibnamefont{and}
  \bibinfo{author}{\bibfnamefont{M.~M.} \bibnamefont{Sheikh-Jabbari}},
  \bibinfo{journal}{Gen. Rel. Grav.} \textbf{\bibinfo{volume}{50}},
  \bibinfo{pages}{110} (\bibinfo{year}{2018}), \eprint{1208.2807}.

\bibitem[{\citenamefont{Adshead
  et~al.}(2013{\natexlab{a}})\citenamefont{Adshead, Martinec, and
  Wyman}}]{Adshead:2013nka}
\bibinfo{author}{\bibfnamefont{P.}~\bibnamefont{Adshead}},
  \bibinfo{author}{\bibfnamefont{E.}~\bibnamefont{Martinec}}, \bibnamefont{and}
  \bibinfo{author}{\bibfnamefont{M.}~\bibnamefont{Wyman}},
  \bibinfo{journal}{JHEP} \textbf{\bibinfo{volume}{09}}, \bibinfo{pages}{087}
  (\bibinfo{year}{2013}{\natexlab{a}}), \eprint{1305.2930}.

\bibitem[{\citenamefont{Adshead
  et~al.}(2016{\natexlab{a}})\citenamefont{Adshead, Martinec, Sfakianakis, and
  Wyman}}]{Adshead:2016omu}
\bibinfo{author}{\bibfnamefont{P.}~\bibnamefont{Adshead}},
  \bibinfo{author}{\bibfnamefont{E.}~\bibnamefont{Martinec}},
  \bibinfo{author}{\bibfnamefont{E.~I.} \bibnamefont{Sfakianakis}},
  \bibnamefont{and} \bibinfo{author}{\bibfnamefont{M.}~\bibnamefont{Wyman}},
  \bibinfo{journal}{JHEP} \textbf{\bibinfo{volume}{12}}, \bibinfo{pages}{137}
  (\bibinfo{year}{2016}{\natexlab{a}}), \eprint{1609.04025}.

\bibitem[{\citenamefont{Adshead and Sfakianakis}(2017)}]{Adshead:2017hnc}
\bibinfo{author}{\bibfnamefont{P.}~\bibnamefont{Adshead}} \bibnamefont{and}
  \bibinfo{author}{\bibfnamefont{E.~I.} \bibnamefont{Sfakianakis}},
  \bibinfo{journal}{JHEP} \textbf{\bibinfo{volume}{08}}, \bibinfo{pages}{130}
  (\bibinfo{year}{2017}), \eprint{1705.03024}.

\bibitem[{\citenamefont{Dimastrogiovanni
  et~al.}(2018)\citenamefont{Dimastrogiovanni, Fasiello, Hardwick, Assadullahi,
  Koyama, and Wands}}]{Dimastrogiovanni:2018xnn}
\bibinfo{author}{\bibfnamefont{E.}~\bibnamefont{Dimastrogiovanni}},
  \bibinfo{author}{\bibfnamefont{M.}~\bibnamefont{Fasiello}},
  \bibinfo{author}{\bibfnamefont{R.~J.} \bibnamefont{Hardwick}},
  \bibinfo{author}{\bibfnamefont{H.}~\bibnamefont{Assadullahi}},
  \bibinfo{author}{\bibfnamefont{K.}~\bibnamefont{Koyama}}, \bibnamefont{and}
  \bibinfo{author}{\bibfnamefont{D.}~\bibnamefont{Wands}},
  \bibinfo{journal}{JCAP} \textbf{\bibinfo{volume}{11}}, \bibinfo{pages}{029}
  (\bibinfo{year}{2018}), \eprint{1806.05474}.

\bibitem[{\citenamefont{Domcke et~al.}(2019{\natexlab{a}})\citenamefont{Domcke,
  Mares, Muia, and Pieroni}}]{Domcke:2018rvv}
\bibinfo{author}{\bibfnamefont{V.}~\bibnamefont{Domcke}},
  \bibinfo{author}{\bibfnamefont{B.}~\bibnamefont{Mares}},
  \bibinfo{author}{\bibfnamefont{F.}~\bibnamefont{Muia}}, \bibnamefont{and}
  \bibinfo{author}{\bibfnamefont{M.}~\bibnamefont{Pieroni}},
  \bibinfo{journal}{JCAP} \textbf{\bibinfo{volume}{04}}, \bibinfo{pages}{034}
  (\bibinfo{year}{2019}{\natexlab{a}}), \eprint{1807.03358}.

\bibitem[{\citenamefont{Domcke et~al.}(2019{\natexlab{b}})\citenamefont{Domcke,
  von Harling, Morgante, and Mukaida}}]{Domcke:2019mnd}
\bibinfo{author}{\bibfnamefont{V.}~\bibnamefont{Domcke}},
  \bibinfo{author}{\bibfnamefont{B.}~\bibnamefont{von Harling}},
  \bibinfo{author}{\bibfnamefont{E.}~\bibnamefont{Morgante}}, \bibnamefont{and}
  \bibinfo{author}{\bibfnamefont{K.}~\bibnamefont{Mukaida}},
  \bibinfo{journal}{JCAP} \textbf{\bibinfo{volume}{10}}, \bibinfo{pages}{032}
  (\bibinfo{year}{2019}{\natexlab{b}}), \eprint{1905.13318}.

\bibitem[{\citenamefont{Adshead et~al.}(2015)\citenamefont{Adshead, Giblin,
  Scully, and Sfakianakis}}]{Adshead:2015pva}
\bibinfo{author}{\bibfnamefont{P.}~\bibnamefont{Adshead}},
  \bibinfo{author}{\bibfnamefont{J.~T.} \bibnamefont{Giblin}},
  \bibinfo{author}{\bibfnamefont{T.~R.} \bibnamefont{Scully}},
  \bibnamefont{and} \bibinfo{author}{\bibfnamefont{E.~I.}
  \bibnamefont{Sfakianakis}}, \bibinfo{journal}{JCAP}
  \textbf{\bibinfo{volume}{12}}, \bibinfo{pages}{034} (\bibinfo{year}{2015}),
  \eprint{1502.06506}.

\bibitem[{\citenamefont{Adshead et~al.}(2017)\citenamefont{Adshead, Giblin, and
  Weiner}}]{Adshead:2017xll}
\bibinfo{author}{\bibfnamefont{P.}~\bibnamefont{Adshead}},
  \bibinfo{author}{\bibfnamefont{J.~T.} \bibnamefont{Giblin}},
  \bibnamefont{and} \bibinfo{author}{\bibfnamefont{Z.~J.}
  \bibnamefont{Weiner}}, \bibinfo{journal}{Phys. Rev. D}
  \textbf{\bibinfo{volume}{96}}, \bibinfo{pages}{123512}
  (\bibinfo{year}{2017}), \eprint{1708.02944}.

\bibitem[{\citenamefont{Adshead et~al.}(2018)\citenamefont{Adshead, Giblin, and
  Weiner}}]{Adshead:2018doq}
\bibinfo{author}{\bibfnamefont{P.}~\bibnamefont{Adshead}},
  \bibinfo{author}{\bibfnamefont{J.~T.} \bibnamefont{Giblin}},
  \bibnamefont{and} \bibinfo{author}{\bibfnamefont{Z.~J.}
  \bibnamefont{Weiner}}, \bibinfo{journal}{Phys. Rev. D}
  \textbf{\bibinfo{volume}{98}}, \bibinfo{pages}{043525}
  (\bibinfo{year}{2018}), \eprint{1805.04550}.

\bibitem[{\citenamefont{Adshead
  et~al.}(2020{\natexlab{a}})\citenamefont{Adshead, Giblin, Pieroni, and
  Weiner}}]{Adshead:2019igv}
\bibinfo{author}{\bibfnamefont{P.}~\bibnamefont{Adshead}},
  \bibinfo{author}{\bibfnamefont{J.~T.} \bibnamefont{Giblin}},
  \bibinfo{author}{\bibfnamefont{M.}~\bibnamefont{Pieroni}}, \bibnamefont{and}
  \bibinfo{author}{\bibfnamefont{Z.~J.} \bibnamefont{Weiner}},
  \bibinfo{journal}{Phys. Rev. Lett.} \textbf{\bibinfo{volume}{124}},
  \bibinfo{pages}{171301} (\bibinfo{year}{2020}{\natexlab{a}}),
  \eprint{1909.12843}.

\bibitem[{\citenamefont{Adshead
  et~al.}(2020{\natexlab{b}})\citenamefont{Adshead, Giblin, Pieroni, and
  Weiner}}]{Adshead:2019lbr}
\bibinfo{author}{\bibfnamefont{P.}~\bibnamefont{Adshead}},
  \bibinfo{author}{\bibfnamefont{J.~T.} \bibnamefont{Giblin}},
  \bibinfo{author}{\bibfnamefont{M.}~\bibnamefont{Pieroni}}, \bibnamefont{and}
  \bibinfo{author}{\bibfnamefont{Z.~J.} \bibnamefont{Weiner}},
  \bibinfo{journal}{Phys. Rev. D} \textbf{\bibinfo{volume}{101}},
  \bibinfo{pages}{083534} (\bibinfo{year}{2020}{\natexlab{b}}),
  \eprint{1909.12842}.

\bibitem[{\citenamefont{Anber and Sorbo}(2012)}]{Anber:2012du}
\bibinfo{author}{\bibfnamefont{M.~M.} \bibnamefont{Anber}} \bibnamefont{and}
  \bibinfo{author}{\bibfnamefont{L.}~\bibnamefont{Sorbo}},
  \bibinfo{journal}{Phys. Rev. D} \textbf{\bibinfo{volume}{85}},
  \bibinfo{pages}{123537} (\bibinfo{year}{2012}), \eprint{1203.5849}.

\bibitem[{\citenamefont{Adshead
  et~al.}(2013{\natexlab{b}})\citenamefont{Adshead, Martinec, and
  Wyman}}]{Adshead:2013qp}
\bibinfo{author}{\bibfnamefont{P.}~\bibnamefont{Adshead}},
  \bibinfo{author}{\bibfnamefont{E.}~\bibnamefont{Martinec}}, \bibnamefont{and}
  \bibinfo{author}{\bibfnamefont{M.}~\bibnamefont{Wyman}},
  \bibinfo{journal}{Phys. Rev. D} \textbf{\bibinfo{volume}{88}},
  \bibinfo{pages}{021302} (\bibinfo{year}{2013}{\natexlab{b}}),
  \eprint{1301.2598}.

\bibitem[{\citenamefont{Namba et~al.}(2013)\citenamefont{Namba,
  Dimastrogiovanni, and Peloso}}]{Namba:2013kia}
\bibinfo{author}{\bibfnamefont{R.}~\bibnamefont{Namba}},
  \bibinfo{author}{\bibfnamefont{E.}~\bibnamefont{Dimastrogiovanni}},
  \bibnamefont{and} \bibinfo{author}{\bibfnamefont{M.}~\bibnamefont{Peloso}},
  \bibinfo{journal}{JCAP} \textbf{\bibinfo{volume}{11}}, \bibinfo{pages}{045}
  (\bibinfo{year}{2013}), \eprint{1308.1366}.

\bibitem[{\citenamefont{Maleknejad}(2016{\natexlab{a}})}]{Maleknejad:2016qjz}
\bibinfo{author}{\bibfnamefont{A.}~\bibnamefont{Maleknejad}},
  \bibinfo{journal}{JHEP} \textbf{\bibinfo{volume}{07}}, \bibinfo{pages}{104}
  (\bibinfo{year}{2016}{\natexlab{a}}), \eprint{1604.03327}.

\bibitem[{\citenamefont{Maleknejad}(2016{\natexlab{b}})}]{Maleknejad:2016dve}
\bibinfo{author}{\bibfnamefont{A.}~\bibnamefont{Maleknejad}}
  (\bibinfo{year}{2016}{\natexlab{b}}), \eprint{1612.05701}.

\bibitem[{\citenamefont{Dimastrogiovanni
  et~al.}(2017)\citenamefont{Dimastrogiovanni, Fasiello, and
  Fujita}}]{Dimastrogiovanni:2016fuu}
\bibinfo{author}{\bibfnamefont{E.}~\bibnamefont{Dimastrogiovanni}},
  \bibinfo{author}{\bibfnamefont{M.}~\bibnamefont{Fasiello}}, \bibnamefont{and}
  \bibinfo{author}{\bibfnamefont{T.}~\bibnamefont{Fujita}},
  \bibinfo{journal}{JCAP} \textbf{\bibinfo{volume}{01}}, \bibinfo{pages}{019}
  (\bibinfo{year}{2017}), \eprint{1608.04216}.

\bibitem[{\citenamefont{Fujita et~al.}(2018)\citenamefont{Fujita, Namba, and
  Tada}}]{Fujita:2017jwq}
\bibinfo{author}{\bibfnamefont{T.}~\bibnamefont{Fujita}},
  \bibinfo{author}{\bibfnamefont{R.}~\bibnamefont{Namba}}, \bibnamefont{and}
  \bibinfo{author}{\bibfnamefont{Y.}~\bibnamefont{Tada}},
  \bibinfo{journal}{Phys. Lett. B} \textbf{\bibinfo{volume}{778}},
  \bibinfo{pages}{17} (\bibinfo{year}{2018}), \eprint{1705.01533}.

\bibitem[{\citenamefont{Caldwell and Devulder}(2018)}]{Caldwell:2017chz}
\bibinfo{author}{\bibfnamefont{R.~R.} \bibnamefont{Caldwell}} \bibnamefont{and}
  \bibinfo{author}{\bibfnamefont{C.}~\bibnamefont{Devulder}},
  \bibinfo{journal}{Phys. Rev. D} \textbf{\bibinfo{volume}{97}},
  \bibinfo{pages}{023532} (\bibinfo{year}{2018}), \eprint{1706.03765}.

\bibitem[{\citenamefont{Agrawal
  et~al.}(2018{\natexlab{a}})\citenamefont{Agrawal, Fujita, and
  Komatsu}}]{Agrawal:2017awz}
\bibinfo{author}{\bibfnamefont{A.}~\bibnamefont{Agrawal}},
  \bibinfo{author}{\bibfnamefont{T.}~\bibnamefont{Fujita}}, \bibnamefont{and}
  \bibinfo{author}{\bibfnamefont{E.}~\bibnamefont{Komatsu}},
  \bibinfo{journal}{Phys. Rev. D} \textbf{\bibinfo{volume}{97}},
  \bibinfo{pages}{103526} (\bibinfo{year}{2018}{\natexlab{a}}),
  \eprint{1707.03023}.

\bibitem[{\citenamefont{Agrawal
  et~al.}(2018{\natexlab{b}})\citenamefont{Agrawal, Fujita, and
  Komatsu}}]{Agrawal:2018mrg}
\bibinfo{author}{\bibfnamefont{A.}~\bibnamefont{Agrawal}},
  \bibinfo{author}{\bibfnamefont{T.}~\bibnamefont{Fujita}}, \bibnamefont{and}
  \bibinfo{author}{\bibfnamefont{E.}~\bibnamefont{Komatsu}},
  \bibinfo{journal}{JCAP} \textbf{\bibinfo{volume}{06}}, \bibinfo{pages}{027}
  (\bibinfo{year}{2018}{\natexlab{b}}), \eprint{1802.09284}.

\bibitem[{\citenamefont{Fujita et~al.}(2019)\citenamefont{Fujita, Sfakianakis,
  and Shiraishi}}]{Fujita:2018ndp}
\bibinfo{author}{\bibfnamefont{T.}~\bibnamefont{Fujita}},
  \bibinfo{author}{\bibfnamefont{E.~I.} \bibnamefont{Sfakianakis}},
  \bibnamefont{and}
  \bibinfo{author}{\bibfnamefont{M.}~\bibnamefont{Shiraishi}},
  \bibinfo{journal}{JCAP} \textbf{\bibinfo{volume}{05}}, \bibinfo{pages}{057}
  (\bibinfo{year}{2019}), \eprint{1812.03667}.

\bibitem[{\citenamefont{Watanabe and Komatsu}(2020)}]{Watanabe:2020ctz}
\bibinfo{author}{\bibfnamefont{Y.}~\bibnamefont{Watanabe}} \bibnamefont{and}
  \bibinfo{author}{\bibfnamefont{E.}~\bibnamefont{Komatsu}}
  (\bibinfo{year}{2020}), \eprint{2004.04350}.

\bibitem[{\citenamefont{Davis et~al.}(2001)\citenamefont{Davis, Dimopoulos,
  Prokopec, and Tornkvist}}]{Davis:2000zp}
\bibinfo{author}{\bibfnamefont{A.-C.} \bibnamefont{Davis}},
  \bibinfo{author}{\bibfnamefont{K.}~\bibnamefont{Dimopoulos}},
  \bibinfo{author}{\bibfnamefont{T.}~\bibnamefont{Prokopec}}, \bibnamefont{and}
  \bibinfo{author}{\bibfnamefont{O.}~\bibnamefont{Tornkvist}},
  \bibinfo{journal}{Phys. Lett. B} \textbf{\bibinfo{volume}{501}},
  \bibinfo{pages}{165} (\bibinfo{year}{2001}), \eprint{astro-ph/0007214}.

\bibitem[{\citenamefont{Dimopoulos et~al.}(2002)\citenamefont{Dimopoulos,
  Prokopec, Tornkvist, and Davis}}]{Dimopoulos:2001wx}
\bibinfo{author}{\bibfnamefont{K.}~\bibnamefont{Dimopoulos}},
  \bibinfo{author}{\bibfnamefont{T.}~\bibnamefont{Prokopec}},
  \bibinfo{author}{\bibfnamefont{O.}~\bibnamefont{Tornkvist}},
  \bibnamefont{and} \bibinfo{author}{\bibfnamefont{A.~C.} \bibnamefont{Davis}},
  \bibinfo{journal}{Phys. Rev. D} \textbf{\bibinfo{volume}{65}},
  \bibinfo{pages}{063505} (\bibinfo{year}{2002}), \eprint{astro-ph/0108093}.

\bibitem[{\citenamefont{Demozzi et~al.}(2009)\citenamefont{Demozzi, Mukhanov,
  and Rubinstein}}]{Demozzi:2009fu}
\bibinfo{author}{\bibfnamefont{V.}~\bibnamefont{Demozzi}},
  \bibinfo{author}{\bibfnamefont{V.}~\bibnamefont{Mukhanov}}, \bibnamefont{and}
  \bibinfo{author}{\bibfnamefont{H.}~\bibnamefont{Rubinstein}},
  \bibinfo{journal}{JCAP} \textbf{\bibinfo{volume}{08}}, \bibinfo{pages}{025}
  (\bibinfo{year}{2009}), \eprint{0907.1030}.

\bibitem[{\citenamefont{Adshead
  et~al.}(2016{\natexlab{b}})\citenamefont{Adshead, Giblin, Scully, and
  Sfakianakis}}]{Adshead:2016iae}
\bibinfo{author}{\bibfnamefont{P.}~\bibnamefont{Adshead}},
  \bibinfo{author}{\bibfnamefont{J.~T.} \bibnamefont{Giblin}},
  \bibinfo{author}{\bibfnamefont{T.~R.} \bibnamefont{Scully}},
  \bibnamefont{and} \bibinfo{author}{\bibfnamefont{E.~I.}
  \bibnamefont{Sfakianakis}}, \bibinfo{journal}{JCAP}
  \textbf{\bibinfo{volume}{10}}, \bibinfo{pages}{039}
  (\bibinfo{year}{2016}{\natexlab{b}}), \eprint{1606.08474}.

\bibitem[{\citenamefont{Kandus et~al.}(2011)\citenamefont{Kandus, Kunze, and
  Tsagas}}]{Kandus:2010nw}
\bibinfo{author}{\bibfnamefont{A.}~\bibnamefont{Kandus}},
  \bibinfo{author}{\bibfnamefont{K.~E.} \bibnamefont{Kunze}}, \bibnamefont{and}
  \bibinfo{author}{\bibfnamefont{C.~G.} \bibnamefont{Tsagas}},
  \bibinfo{journal}{Phys. Rept.} \textbf{\bibinfo{volume}{505}},
  \bibinfo{pages}{1} (\bibinfo{year}{2011}), \eprint{1007.3891}.

\bibitem[{\citenamefont{Cui et~al.}(2018)\citenamefont{Cui, Lewicki, Morrissey,
  and Wells}}]{Cui:2017ufi}
\bibinfo{author}{\bibfnamefont{Y.}~\bibnamefont{Cui}},
  \bibinfo{author}{\bibfnamefont{M.}~\bibnamefont{Lewicki}},
  \bibinfo{author}{\bibfnamefont{D.~E.} \bibnamefont{Morrissey}},
  \bibnamefont{and} \bibinfo{author}{\bibfnamefont{J.~D.} \bibnamefont{Wells}},
  \bibinfo{journal}{Phys. Rev. D} \textbf{\bibinfo{volume}{97}},
  \bibinfo{pages}{123505} (\bibinfo{year}{2018}), \eprint{1711.03104}.

\bibitem[{\citenamefont{Allahverdi et~al.}(2021)}]{Allahverdi:2020bys}
\bibinfo{author}{\bibfnamefont{R.}~\bibnamefont{Allahverdi}}
  \bibnamefont{et~al.}, \bibinfo{journal}{Open J. Astrophys.}
  \textbf{\bibinfo{volume}{4}} (\bibinfo{year}{2021}), \eprint{2006.16182}.

\bibitem[{\citenamefont{Seto and Yokoyama}(2003)}]{Seto:2003kc}
\bibinfo{author}{\bibfnamefont{N.}~\bibnamefont{Seto}} \bibnamefont{and}
  \bibinfo{author}{\bibfnamefont{J.}~\bibnamefont{Yokoyama}},
  \bibinfo{journal}{J. Phys. Soc. Jap.} \textbf{\bibinfo{volume}{72}},
  \bibinfo{pages}{3082} (\bibinfo{year}{2003}), \eprint{gr-qc/0305096}.

\bibitem[{\citenamefont{Boyle and Steinhardt}(2008)}]{Boyle:2005se}
\bibinfo{author}{\bibfnamefont{L.~A.} \bibnamefont{Boyle}} \bibnamefont{and}
  \bibinfo{author}{\bibfnamefont{P.~J.} \bibnamefont{Steinhardt}},
  \bibinfo{journal}{Phys. Rev. D} \textbf{\bibinfo{volume}{77}},
  \bibinfo{pages}{063504} (\bibinfo{year}{2008}), \eprint{astro-ph/0512014}.

\bibitem[{\citenamefont{Watanabe and Komatsu}(2006)}]{Watanabe:2006qe}
\bibinfo{author}{\bibfnamefont{Y.}~\bibnamefont{Watanabe}} \bibnamefont{and}
  \bibinfo{author}{\bibfnamefont{E.}~\bibnamefont{Komatsu}},
  \bibinfo{journal}{Phys. Rev. D} \textbf{\bibinfo{volume}{73}},
  \bibinfo{pages}{123515} (\bibinfo{year}{2006}), \eprint{astro-ph/0604176}.

\bibitem[{\citenamefont{Nakayama et~al.}(2008)\citenamefont{Nakayama, Saito,
  Suwa, and Yokoyama}}]{Nakayama:2008ip}
\bibinfo{author}{\bibfnamefont{K.}~\bibnamefont{Nakayama}},
  \bibinfo{author}{\bibfnamefont{S.}~\bibnamefont{Saito}},
  \bibinfo{author}{\bibfnamefont{Y.}~\bibnamefont{Suwa}}, \bibnamefont{and}
  \bibinfo{author}{\bibfnamefont{J.}~\bibnamefont{Yokoyama}},
  \bibinfo{journal}{Phys. Rev. D} \textbf{\bibinfo{volume}{77}},
  \bibinfo{pages}{124001} (\bibinfo{year}{2008}), \eprint{0802.2452}.

\bibitem[{\citenamefont{Kuroyanagi et~al.}(2011)\citenamefont{Kuroyanagi,
  Nakayama, and Saito}}]{Kuroyanagi:2011fy}
\bibinfo{author}{\bibfnamefont{S.}~\bibnamefont{Kuroyanagi}},
  \bibinfo{author}{\bibfnamefont{K.}~\bibnamefont{Nakayama}}, \bibnamefont{and}
  \bibinfo{author}{\bibfnamefont{S.}~\bibnamefont{Saito}},
  \bibinfo{journal}{Phys. Rev. D} \textbf{\bibinfo{volume}{84}},
  \bibinfo{pages}{123513} (\bibinfo{year}{2011}), \eprint{1110.4169}.

\bibitem[{\citenamefont{Saikawa and Shirai}(2018)}]{Saikawa:2018rcs}
\bibinfo{author}{\bibfnamefont{K.}~\bibnamefont{Saikawa}} \bibnamefont{and}
  \bibinfo{author}{\bibfnamefont{S.}~\bibnamefont{Shirai}},
  \bibinfo{journal}{JCAP} \textbf{\bibinfo{volume}{05}}, \bibinfo{pages}{035}
  (\bibinfo{year}{2018}), \eprint{1803.01038}.

\bibitem[{\citenamefont{Caldwell et~al.}(2019)\citenamefont{Caldwell, Smith,
  and Walker}}]{Caldwell:2018giq}
\bibinfo{author}{\bibfnamefont{R.~R.} \bibnamefont{Caldwell}},
  \bibinfo{author}{\bibfnamefont{T.~L.} \bibnamefont{Smith}}, \bibnamefont{and}
  \bibinfo{author}{\bibfnamefont{D.~G.~E.} \bibnamefont{Walker}},
  \bibinfo{journal}{Phys. Rev. D} \textbf{\bibinfo{volume}{100}},
  \bibinfo{pages}{043513} (\bibinfo{year}{2019}), \eprint{1812.07577}.

\bibitem[{\citenamefont{Figueroa and Tanin}(2019)}]{Figueroa:2019paj}
\bibinfo{author}{\bibfnamefont{D.~G.} \bibnamefont{Figueroa}} \bibnamefont{and}
  \bibinfo{author}{\bibfnamefont{E.~H.} \bibnamefont{Tanin}},
  \bibinfo{journal}{JCAP} \textbf{\bibinfo{volume}{08}}, \bibinfo{pages}{011}
  (\bibinfo{year}{2019}), \eprint{1905.11960}.

\bibitem[{\citenamefont{Bielefeld and Caldwell}(2015)}]{Bielefeld:2015daa}
\bibinfo{author}{\bibfnamefont{J.}~\bibnamefont{Bielefeld}} \bibnamefont{and}
  \bibinfo{author}{\bibfnamefont{R.~R.} \bibnamefont{Caldwell}},
  \bibinfo{journal}{Phys. Rev. D} \textbf{\bibinfo{volume}{91}},
  \bibinfo{pages}{124004} (\bibinfo{year}{2015}), \eprint{1503.05222}.

\bibitem[{\citenamefont{Aghanim et~al.}(2020)}]{Aghanim:2018eyx}
\bibinfo{author}{\bibfnamefont{N.}~\bibnamefont{Aghanim}} \bibnamefont{et~al.}
  (\bibinfo{collaboration}{Planck}), \bibinfo{journal}{Astron. Astrophys.}
  \textbf{\bibinfo{volume}{641}}, \bibinfo{pages}{A6} (\bibinfo{year}{2020}),
  \eprint{1807.06209}.

\bibitem[{\citenamefont{Fields et~al.}(2020)\citenamefont{Fields, Olive, Yeh,
  and Young}}]{Fields:2019pfx}
\bibinfo{author}{\bibfnamefont{B.~D.} \bibnamefont{Fields}},
  \bibinfo{author}{\bibfnamefont{K.~A.} \bibnamefont{Olive}},
  \bibinfo{author}{\bibfnamefont{T.-H.} \bibnamefont{Yeh}}, \bibnamefont{and}
  \bibinfo{author}{\bibfnamefont{C.}~\bibnamefont{Young}},
  \bibinfo{journal}{JCAP} \textbf{\bibinfo{volume}{03}}, \bibinfo{pages}{010}
  (\bibinfo{year}{2020}), \bibinfo{note}{[Erratum: JCAP 11, E02 (2020)]},
  \eprint{1912.01132}.

\bibitem[{\citenamefont{Gertsenshteyn}(1962)}]{Gertsenshteyn1962}
\bibinfo{author}{\bibfnamefont{M.~E.} \bibnamefont{Gertsenshteyn}},
  \bibinfo{journal}{Sov. Phys. JETP} \textbf{\bibinfo{volume}{14}},
  \bibinfo{pages}{84} (\bibinfo{year}{1962}).

\bibitem[{\citenamefont{Poznanin}(1969)}]{Poznanin1969}
\bibinfo{author}{\bibfnamefont{P.-L.} \bibnamefont{Poznanin}},
  \bibinfo{journal}{Sov. Phys. J.} \textbf{\bibinfo{volume}{12}},
  \bibinfo{pages}{1296} (\bibinfo{year}{1969}).

\bibitem[{\citenamefont{Boccaletti et~al.}(1970)\citenamefont{Boccaletti,
  Sabbata, Fortini, and Gualdi}}]{Boccaletti1970}
\bibinfo{author}{\bibfnamefont{D.}~\bibnamefont{Boccaletti}},
  \bibinfo{author}{\bibfnamefont{V.}~\bibnamefont{Sabbata}},
  \bibinfo{author}{\bibfnamefont{P.}~\bibnamefont{Fortini}}, \bibnamefont{and}
  \bibinfo{author}{\bibfnamefont{C.}~\bibnamefont{Gualdi}},
  \bibinfo{journal}{Nuovo Cimento V Serie} \textbf{\bibinfo{volume}{70}},
  \bibinfo{pages}{129} (\bibinfo{year}{1970}).

\bibitem[{\citenamefont{Zeldovich}(1974)}]{Zeldovich1974}
\bibinfo{author}{\bibfnamefont{Y.~B.} \bibnamefont{Zeldovich}},
  \bibinfo{journal}{Sov. Phys. JETP} \textbf{\bibinfo{volume}{38}},
  \bibinfo{pages}{652} (\bibinfo{year}{1974}).

\bibitem[{\citenamefont{Caldwell et~al.}(2016)\citenamefont{Caldwell, Devulder,
  and Maksimova}}]{Caldwell:2016sut}
\bibinfo{author}{\bibfnamefont{R.~R.} \bibnamefont{Caldwell}},
  \bibinfo{author}{\bibfnamefont{C.}~\bibnamefont{Devulder}}, \bibnamefont{and}
  \bibinfo{author}{\bibfnamefont{N.~A.} \bibnamefont{Maksimova}},
  \bibinfo{journal}{Phys. Rev. D} \textbf{\bibinfo{volume}{94}},
  \bibinfo{pages}{063005} (\bibinfo{year}{2016}), \eprint{1604.08939}.

\bibitem[{\citenamefont{Gluscevic and Kamionkowski}(2010)}]{Gluscevic:2010vv}
\bibinfo{author}{\bibfnamefont{V.}~\bibnamefont{Gluscevic}} \bibnamefont{and}
  \bibinfo{author}{\bibfnamefont{M.}~\bibnamefont{Kamionkowski}},
  \bibinfo{journal}{Phys. Rev. D} \textbf{\bibinfo{volume}{81}},
  \bibinfo{pages}{123529} (\bibinfo{year}{2010}), \eprint{1002.1308}.

\bibitem[{\citenamefont{Gerbino et~al.}(2016)\citenamefont{Gerbino, Gruppuso,
  Natoli, Shiraishi, and Melchiorri}}]{Gerbino:2016mqb}
\bibinfo{author}{\bibfnamefont{M.}~\bibnamefont{Gerbino}},
  \bibinfo{author}{\bibfnamefont{A.}~\bibnamefont{Gruppuso}},
  \bibinfo{author}{\bibfnamefont{P.}~\bibnamefont{Natoli}},
  \bibinfo{author}{\bibfnamefont{M.}~\bibnamefont{Shiraishi}},
  \bibnamefont{and}
  \bibinfo{author}{\bibfnamefont{A.}~\bibnamefont{Melchiorri}},
  \bibinfo{journal}{JCAP} \textbf{\bibinfo{volume}{07}}, \bibinfo{pages}{044}
  (\bibinfo{year}{2016}), \eprint{1605.09357}.

\bibitem[{\citenamefont{Thorne et~al.}(2018)\citenamefont{Thorne, Fujita,
  Hazumi, Katayama, Komatsu, and Shiraishi}}]{Thorne:2017jft}
\bibinfo{author}{\bibfnamefont{B.}~\bibnamefont{Thorne}},
  \bibinfo{author}{\bibfnamefont{T.}~\bibnamefont{Fujita}},
  \bibinfo{author}{\bibfnamefont{M.}~\bibnamefont{Hazumi}},
  \bibinfo{author}{\bibfnamefont{N.}~\bibnamefont{Katayama}},
  \bibinfo{author}{\bibfnamefont{E.}~\bibnamefont{Komatsu}}, \bibnamefont{and}
  \bibinfo{author}{\bibfnamefont{M.}~\bibnamefont{Shiraishi}},
  \bibinfo{journal}{Phys. Rev. D} \textbf{\bibinfo{volume}{97}},
  \bibinfo{pages}{043506} (\bibinfo{year}{2018}), \eprint{1707.03240}.

\bibitem[{\citenamefont{Thrane and Romano}(2013)}]{Thrane:2013oya}
\bibinfo{author}{\bibfnamefont{E.}~\bibnamefont{Thrane}} \bibnamefont{and}
  \bibinfo{author}{\bibfnamefont{J.~D.} \bibnamefont{Romano}},
  \bibinfo{journal}{Phys. Rev. D} \textbf{\bibinfo{volume}{88}},
  \bibinfo{pages}{124032} (\bibinfo{year}{2013}), \eprint{1310.5300}.

\bibitem[{\citenamefont{Smith and Caldwell}(2019)}]{Smith:2019wny}
\bibinfo{author}{\bibfnamefont{T.~L.} \bibnamefont{Smith}} \bibnamefont{and}
  \bibinfo{author}{\bibfnamefont{R.}~\bibnamefont{Caldwell}},
  \bibinfo{journal}{Phys. Rev. D} \textbf{\bibinfo{volume}{100}},
  \bibinfo{pages}{104055} (\bibinfo{year}{2019}), \eprint{1908.00546}.

\bibitem[{\citenamefont{Schmitz}(2021)}]{Schmitz:2020syl}
\bibinfo{author}{\bibfnamefont{K.}~\bibnamefont{Schmitz}},
  \bibinfo{journal}{JHEP} \textbf{\bibinfo{volume}{01}}, \bibinfo{pages}{097}
  (\bibinfo{year}{2021}), \eprint{2002.04615}.

\bibitem[{\citenamefont{Amaro-Seoane et~al.}(2017)}]{Audley:2017drz}
\bibinfo{author}{\bibfnamefont{P.}~\bibnamefont{Amaro-Seoane}}
  \bibnamefont{et~al.} (\bibinfo{collaboration}{LISA}) (\bibinfo{year}{2017}),
  \eprint{1702.00786}.

\bibitem[{\citenamefont{Abbott et~al.}(2017)}]{Evans:2016mbw}
\bibinfo{author}{\bibfnamefont{B.~P.} \bibnamefont{Abbott}}
  \bibnamefont{et~al.} (\bibinfo{collaboration}{LIGO Scientific}),
  \bibinfo{journal}{Class. Quant. Grav.} \textbf{\bibinfo{volume}{34}},
  \bibinfo{pages}{044001} (\bibinfo{year}{2017}), \bibinfo{note}{\url
  {https://dcc.ligo.org/LIGO-P1600143/public}}, \eprint{1607.08697}.

\bibitem[{\citenamefont{Hild et~al.}(2011)}]{Hild:2010id}
\bibinfo{author}{\bibfnamefont{S.}~\bibnamefont{Hild}} \bibnamefont{et~al.},
  \bibinfo{journal}{Class. Quant. Grav.} \textbf{\bibinfo{volume}{28}},
  \bibinfo{pages}{094013} (\bibinfo{year}{2011}), \eprint{1012.0908}.

\bibitem[{\citenamefont{Punturo}()}]{Punturo:2021ETsensitivity}
\bibinfo{author}{\bibfnamefont{M.}~\bibnamefont{Punturo}},
  \emph{\bibinfo{title}{{ET} sensitivities page}},
  \bibinfo{howpublished}{\url{http://www.et-gw.eu/index.php/etsensitivities}}.

\bibitem[{\citenamefont{Crowder and Cornish}(2005)}]{Crowder:2005nr}
\bibinfo{author}{\bibfnamefont{J.}~\bibnamefont{Crowder}} \bibnamefont{and}
  \bibinfo{author}{\bibfnamefont{N.~J.} \bibnamefont{Cornish}},
  \bibinfo{journal}{Phys. Rev. D} \textbf{\bibinfo{volume}{72}},
  \bibinfo{pages}{083005} (\bibinfo{year}{2005}), \eprint{gr-qc/0506015}.

\bibitem[{\citenamefont{Harry et~al.}(2006)\citenamefont{Harry, Fritschel,
  Shaddock, Folkner, and Phinney}}]{Harry:2006fi}
\bibinfo{author}{\bibfnamefont{G.~M.} \bibnamefont{Harry}},
  \bibinfo{author}{\bibfnamefont{P.}~\bibnamefont{Fritschel}},
  \bibinfo{author}{\bibfnamefont{D.~A.} \bibnamefont{Shaddock}},
  \bibinfo{author}{\bibfnamefont{W.}~\bibnamefont{Folkner}}, \bibnamefont{and}
  \bibinfo{author}{\bibfnamefont{E.~S.} \bibnamefont{Phinney}},
  \bibinfo{journal}{Class. Quant. Grav.} \textbf{\bibinfo{volume}{23}},
  \bibinfo{pages}{4887} (\bibinfo{year}{2006}), \bibinfo{note}{[Erratum:
  Class.Quant.Grav. 23, 7361 (2006)]}.

\bibitem[{\citenamefont{Maleknejad and
  Sheikh-Jabbari}(2011)}]{Maleknejad:2011sq}
\bibinfo{author}{\bibfnamefont{A.}~\bibnamefont{Maleknejad}} \bibnamefont{and}
  \bibinfo{author}{\bibfnamefont{M.~M.} \bibnamefont{Sheikh-Jabbari}},
  \bibinfo{journal}{Phys. Rev. D} \textbf{\bibinfo{volume}{84}},
  \bibinfo{pages}{043515} (\bibinfo{year}{2011}), \eprint{1102.1932}.

\bibitem[{\citenamefont{Ma and Bertschinger}(1995)}]{Ma:1995ey}
\bibinfo{author}{\bibfnamefont{C.-P.} \bibnamefont{Ma}} \bibnamefont{and}
  \bibinfo{author}{\bibfnamefont{E.}~\bibnamefont{Bertschinger}},
  \bibinfo{journal}{Astrophys. J.} \textbf{\bibinfo{volume}{455}},
  \bibinfo{pages}{7} (\bibinfo{year}{1995}), \eprint{astro-ph/9506072}.

\bibitem[{\citenamefont{Mukhanov et~al.}(1992)\citenamefont{Mukhanov, Feldman,
  and Brandenberger}}]{Mukhanov:1990me}
\bibinfo{author}{\bibfnamefont{V.~F.} \bibnamefont{Mukhanov}},
  \bibinfo{author}{\bibfnamefont{H.~A.} \bibnamefont{Feldman}},
  \bibnamefont{and} \bibinfo{author}{\bibfnamefont{R.~H.}
  \bibnamefont{Brandenberger}}, \bibinfo{journal}{Phys. Rept.}
  \textbf{\bibinfo{volume}{215}}, \bibinfo{pages}{203} (\bibinfo{year}{1992}).

\end{thebibliography}

\end{document}